\documentclass[twocolumn]{aastex61}

\usepackage{textcomp}
\usepackage{array,multirow}
\usepackage{natbib}
\usepackage{amsmath}
\usepackage{gensymb}
\usepackage[maxfloats=254]{morefloats}
\begin{document}

\title{Extended X-ray Emission in Compton Thick AGN with Deep \textit{Chandra} Observations}

\author{Mackenzie L. Jones}
\affil{Center for Astrophysics | Harvard \& Smithsonian, 80 Garden St, Cambridge, MA 02138, USA}

\author{Kieran Parker}
\affil{Physics and Astronomy, University of Southampton, Highfield, SO17 1BJ, UK}
\affil{Center for Astrophysics | Harvard \& Smithsonian, 80 Garden St, Cambridge, MA 02138, USA}

\author{G. Fabbiano}
\affil{Center for Astrophysics | Harvard \& Smithsonian, 80 Garden St, Cambridge, MA 02138, USA}

\author{Martin Elvis}
\affil{Center for Astrophysics | Harvard \& Smithsonian, 80 Garden St, Cambridge, MA 02138, USA}

\author{W. P. Maksym}
\affil{Center for Astrophysics | Harvard \& Smithsonian, 80 Garden St, Cambridge, MA 02138, USA}

\author{A. Paggi}
\affil{INAF-Osservatorio Astrofisica di Torino, Via Osservatorio 20, 10025 Pino Torinese, Italy}

\author{Jingzhe Ma}
\affil{Center for Astrophysics | Harvard \& Smithsonian, 80 Garden St, Cambridge, MA 02138, USA}

\author{M. Karovska}
\affil{Center for Astrophysics $\vert$ Harvard \& Smithsonian, 60 Garden St, Cambridge, MA 02138, USA}

\author{A. Siemiginowska}
\affil{Center for Astrophysics $\vert$ Harvard \& Smithsonian, 60 Garden St, Cambridge, MA 02138, USA}

\author{Junfeng Wang}
\affil{Department of Astronomy, Xiamen University, Xiamen, 361005, China}

%%%%%%%%%%%%%%%%%%%%%%%%%%%%%%%%%%%%%%%%%%%%%%%%%
% ABSTRACT
%%%%%%%%%%%%%%%%%%%%%%%%%%%%%%%%%%%%%%%%%%%%%%%%%
\begin{abstract}

We present the spatial analysis of five Compton thick (CT) active galactic nuclei (AGNs), including MKN 573, NGC 1386, NGC 3393, NGC 5643, and NGC 7212, for which high resolution \textit{Chandra} observations are available. For each source, we find hard X-ray emission ($>$3 keV) extending to $\sim$kpc scales along the ionization cone, and for some sources, in the cross-cone region. This collection represents the first, high-signal sample of CT AGN with extended hard X-ray emission for which we can begin to build a more complete picture of this new population of AGN.
We investigate the energy dependence of the extended X-ray emission, including possible dependencies on host galaxy and AGN properties, and find a correlation between the excess emission and obscuration, suggesting a connection between the nuclear obscuring material and the galactic molecular clouds. Furthermore, we find that the soft X-ray emission extends farther than the hard X-rays along the ionization cone, which may be explained by a galactocentric radial dependence on the density of molecular clouds due to the orientation of the ionization cone with respect to the galactic disk. These results are consistent with other CT AGN with observed extended hard X-ray emission (e.g., ESO 428-G014 and the \citealt{Ma20} CT AGN sample), further demonstrating the ubiquity of extended hard X-ray emission in CT AGN.

\end{abstract}


%%%%%%%%%%%%%%%%%%%%%%%%%%%%%%%%%%%%%%%%%%%%%%%%%
% INTRODUCTION
%%%%%%%%%%%%%%%%%%%%%%%%%%%%%%%%%%%%%%%%%%%%%%%%%
\section{Introduction}\label{sec:intro}

Recent \textit{Chandra} observations of nearby, Compton thick (CT) active galactic nuclei (AGNs) have uncovered kiloparsec-scale extended hard X-ray and Fe K$\alpha$ line emission regions that have challenged our understanding of the origin and extent of high energy photons and what impact they may have on their host galaxies (e.g., \textit{Circinus}, \citealt{Are14}; NGC 1068, \citealt{Bau15}; ESO 428-G014, \citealt{Fab17}; NGC 7212, \citealt{Jon20}). 

In the classical picture, the hard X-ray continuum and fluorescent Fe K lines that are observed in AGN are generated by the excitation of the obscuring material in the inner parsecs. Observing this characteristic energetic emission on host galaxy scales is unexpected. The presence of extended emission outside of this inner region has interesting consequences for AGN feedback and its impact on the surrounding medium.

The first well-studied case of hard X-ray emission observed outside of the nuclear region was in the CT AGN, ESO 428-G014 (\citealt{Fab17,Fab18II,Fab18III,Fab19}). ESO 428-G014 exhibits extended emission predominately in the soft X-ray band, but also has significant extent in the hard X-rays, including the band around Fe K$\alpha$ ($6.1-6.5$ keV). The spectrum is best described by a complex mixture of thermal and photoionization models that are consistent with the picture of energetic emission extending from the nucleus into the host-galaxy. 
Since this discovery, a handful of other CT AGN have been individually identified as having extended emission on $\sim$kpc scales, including NGC 7212. NGC 7212 is the farthest of these sources examined thus far ($z=0.0266$), and has a spectrum that is best described by a complex mixture of physical models, as in ESO 428-G014 (\citealt{Jon20}). 

For CT AGNs, in particular, energetic X-ray photons in the inner regions are expected to be completely attenuated by an optically thick, molecular dust torus-like structure that only allows radiation to propagate out along the torus opening angle as an ionization cone (e.g., \citealt{Urr95,Net15}). 
However, in the case of ESO 428-G014, significant, extended emission is found in the ``cross-cone'' region, aligned with this CT torus (\citealt{Fab18II}). Similar significant features are also found in NGC 7212 (\citealt{Jon20}). This suggests that rather than acting as a homogenous screen in this ``cross-cone'' direction, the torus is likely porous, allowing these highly energetic photons to ``leak'' out and interact with the surrounding medium.

A question remains then about how ubiquitous the extended hard X-ray emission is in CT AGN. To explore this further, \citet{Ma20} collected \textit{Chandra} observations for seven CT AGN, not previously known to have extended emission and compared them with ESO 428-G014 and NGC 7212. They demonstrate that the extended hard X-ray emission, including that from Fe K$\alpha$, is a characteristic feature of these obscured sources. Furthermore, this emission can contribute between $\sim8-36$ \% of the total observed emission in $3-7$ keV. This sample, however, is limited by low number counts that make probing regional dependencies (i.e., cone versus cross-cone) challenging.

In this paper we investigate the energy dependence of the hard X-ray emission extent, including dependencies on host galaxy and AGN properties, using a sample of four CT AGN: MKN 573, NGC 1386, NGC 3393, and NGC 5643, to compare with NGC 7212, and the \citet{Ma20} sample. 
The \textit{Chandra} observations and data reduction for these CT AGN are described in Section \ref{sec:obs}. For each source we report on the spatial extent of the X-ray emission in Section \ref{sec:rp}, and discuss the implications of this population of extended X-ray AGN in Section \ref{sec:dis}. Our findings and conclusions are summarized in Section \ref{sec:con}.

%%%%%%%%%%%%%%%%%%%%%%%%%%%%%%%%%%%%%%%%%%%%%%%%%
% OBSERVATIONS & ANALYSIS
%%%%%%%%%%%%%%%%%%%%%%%%%%%%%%%%%%%%%%%%%%%%%%%%%
\section{Observations}\label{sec:obs}

%%%%%%%%%%%%%%%%%%%%%
% TABLE : OBSERVATION SUMMARY
\begin{table}
\centering
\caption{\textit{Chandra} ACIS-S Observation Log} \label{tab:obs}
\begin{footnotesize}
\begin{tabular}{llllll}
\hline\hline
Source & ObsID & t$_{\text{exp}}$ (ks) & PI & Date \\
\hline
MKN 573 & 7745 & 38.08 & Bianchi & 2006-11-18 \\
 & 12294 & 9.92 & Wang & 2010-09-16 \\
 & 13124 & 52.37 & Wang & 2010-09-17 \\
 & 13125 & 16.83 & Wang & 2010-09-19 \\ \hline
NGC 1386 & 4076 & 19.64 & Kraemer & 2003-11-19 \\
 & 12289 & 17.32 & Wang & 2011-04-13 \\
 & 13185 & 29.67 & Wang & 2011-04-13 \\
 & 13257 & 33.82 & Wang & 2011-04-14 \\ \hline
NGC 3393 & 4868 & 29.33 & Levenson & 2004-02-28 \\
 & 12290 & 69.16 & Wang & 2011-03-12 \\
 & 20496 & 48.25 & Maksym & 2019-04-04 \\
 & 20497 & 39.54 & Maksym & 2018-03-19 \\
 & 20498 & 44.52 & Maksym & 2018-03-18 \\
 & 21039 & 44.47 & Maksym & 2019-04-09 \\
 & 21047 & 95.84 & Maksym & 2018-03-23 \\
 & 21048 & 40.43 & Maksym & 2018-07-23 \\
 & 22077 & 47.15 & Maksym & 2019-03-11 \\
 & 22078 & 79.06 & Maksym & 2019-08-06 \\ \hline
NGC 5643 & 17031 & 72.12 & Fabbiano & 2015-05-21 \\
 & 17664 & 41.53 & Fabbiano & 2015-12-26 \\ \hline
NGC 7212 & 4078 & 19.9 & Kraemer & 2003-07-22 \\
 & 20372 & 49.42 & Fabbiano & 2018-08-08 \\
 & 21668 & 51.38 & Fabbiano & 2018-08-11 \\
 & 21672 & 27.21 & Fabbiano & 2018-09-07 \\
\hline
\end{tabular}
\end{footnotesize}
\end{table}
%%%%%%%%%%%%%%%%%%%%%

%%%%%%%%%%%%%%%%%%%%%
% TEXT
%%%%%%%%%%%%%%%%%%%%%
Our sample consists of five CT AGN ($\log N_H > 24$ cm$^{-2}$) with archival \textit{Chandra} aim point observations (Table \ref{tab:obs}). These observations were reprocessed with \textit{Chandra-repro} and analyzed using CIAO 4.11 (\citealt{Fru06}) and CALDB 4.8.2 and inspected for high background flares ($>$$3\sigma$). Each individual observation for each source was exposure-corrected and merged\footnote{http://cxc.harvard.edu/ciao/threads/combine/},\footnote{http://cxc.harvard.edu/ciao/threads/merge\_all/}. 
We visually inspected each observation using the CIAO image analysis tools available in SAOImage ds9\footnote{http://ds9.si.edu} and enabled $1/8$ native ACIS-S sub-pixel binning ($0.062$\arcsec) to improve the spatial resolution of these observations (as in e.g., \citealt{Tsu01,Wan11}). 
From the merged observations for each source, we generated a full band (0.3-8.0 keV) adaptively smoothed image (using \textit{dmimgadapt} from the ds9 CIAO package\footnote{http://cxc.harvard.edu/ciao/gallery/smooth.html}) to investigate the detailed morphology of our CT AGN. The smoothing parameters used in this analysis were chosen to highlight the extended emission: $0.5-15$ pixel scale with 5 counts under the gaussian kernel for 30 iterations, unless otherwise indicated.

%%%%%%%%%%%%%%%%%%%%%%%%
\subsection{MKN 573}
%%%%%%%%%%%%%%%%%%%%%%%%

%%%%%%%%%%%%%%%%%%%%%
% FIGURE : FULL IMAGE
\begin{figure}
\begin{center}
\resizebox{75mm}{!}{\includegraphics{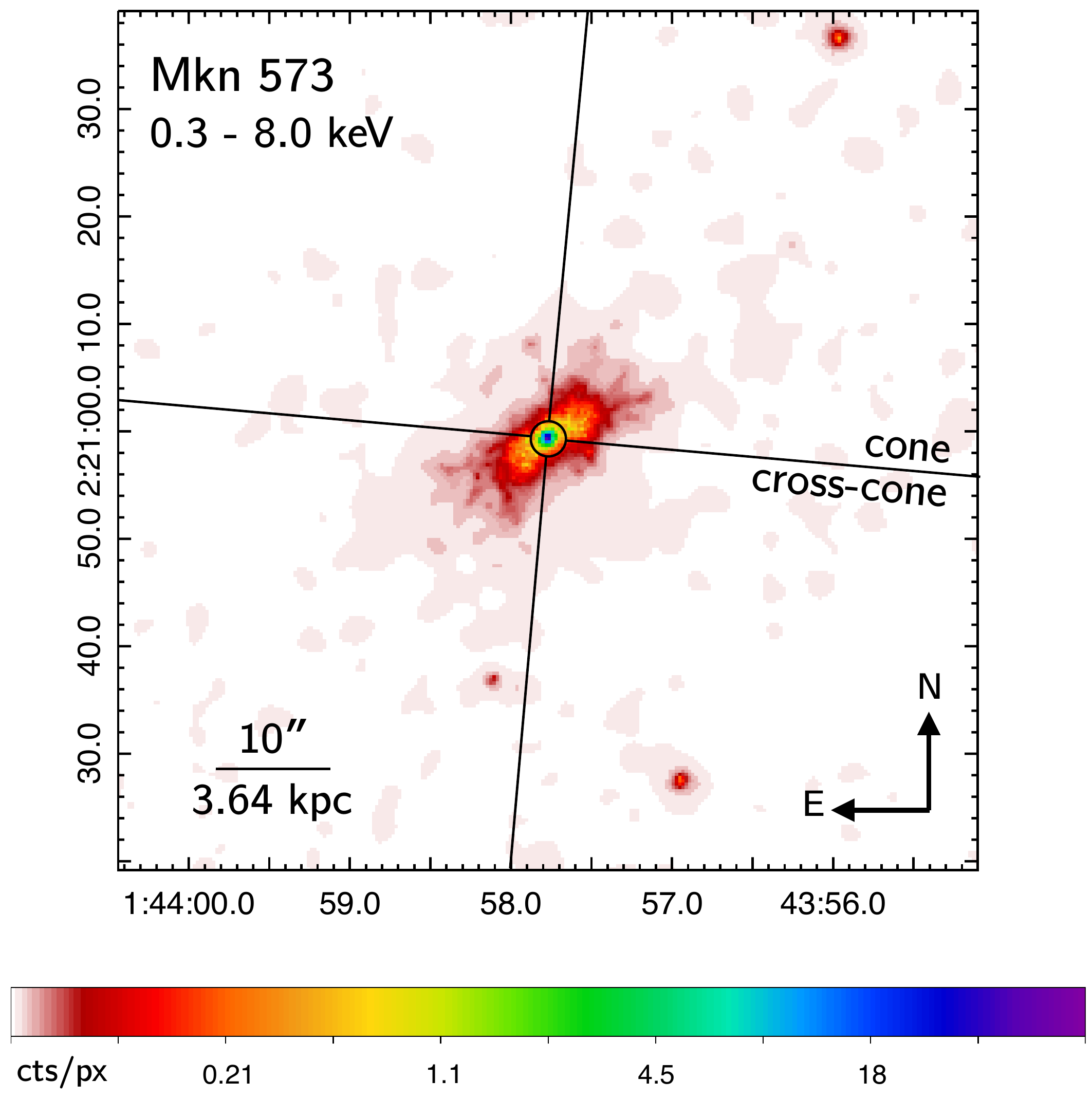}} \\
\caption{Merged $0.3-8.0$ keV \textit{Chandra} ACIS image of MKN 573 with applied adaptive gaussian smoothing (\textit{dmingadapt}; $0.5-15$ pixel scales, 5 counts under kernel, 30 iterations) on image pixel $=1/8$ ACIS pixel. The image contours are logarithmic with colors corresponding to the number of counts per image pixel. The box size is 80\arcsec\, $\times$ 80\arcsec (29.12 $\times$ 29.12 kpc). Also shown are the 1.5\arcsec\,(0.546 kpc) circular region and cone/cross-cone quadrants used in our analysis of the X-ray extent. \label{fig:img:573}}
\end{center}
\end{figure}
%%%%%%%%%%%%%%%%%%%%%

%%%%%%%%%%%%%%%%%%%%%
% TEXT
%%%%%%%%%%%%%%%%%%%%%
MKN 573 is an SAB0 type galaxy at RA=01:43:57.80 (25.991\degree), Dec=+02:20:59.65 (2.350\degree),
and $z=0.0172$ (D$_{\text{lum}}\sim72$ Mpc), with a double radio source (\citealt{Nag99}). The AGN ($M_{BH}=2\times10^{7}$ M$\odot$, \citealt{Bia07}; $L_{x,2-10 keV}=2.2\times10^{43}$ erg s$^{-1}$, \citealt{Ram09}) is optically classified as a Seyfert 2, but \citet{Ram08} finds evidence that a narrow-line Seyfert 1 is hiding beneath the CT obscuring material ($N_H >1.6\times10^{24}$ cm$^{-2}$; \citealt{Gua05}). MKN 573 has been previously shown to have extended, biconical soft X-ray emission on kpc scales (\citealt{Gon10,Pag12}). We extend this analysis to the harder X-rays, focusing specifically on the spatial extent of the 6 to 7 keV band where we expect to find Fe K$\alpha$ line emission. The full-band ($0.3-8.0$ keV) adaptively smoothed image of MKN 573 is shown in Figure \ref{fig:img:573}. Consistent with previous X-ray observations, we observe biconical emission that we separate into ionization cone/cross-cone regions (as indicated in Figure \ref{fig:img:573}).

%%%%%%%%%%%%%%%%%%%%%%%%
\subsection{NGC 1386}
%%%%%%%%%%%%%%%%%%%%%%%%

%%%%%%%%%%%%%%%%%%%%%
% FIGURE : FULL IMAGE
\begin{figure}
\begin{center}
\resizebox{75mm}{!}{\includegraphics{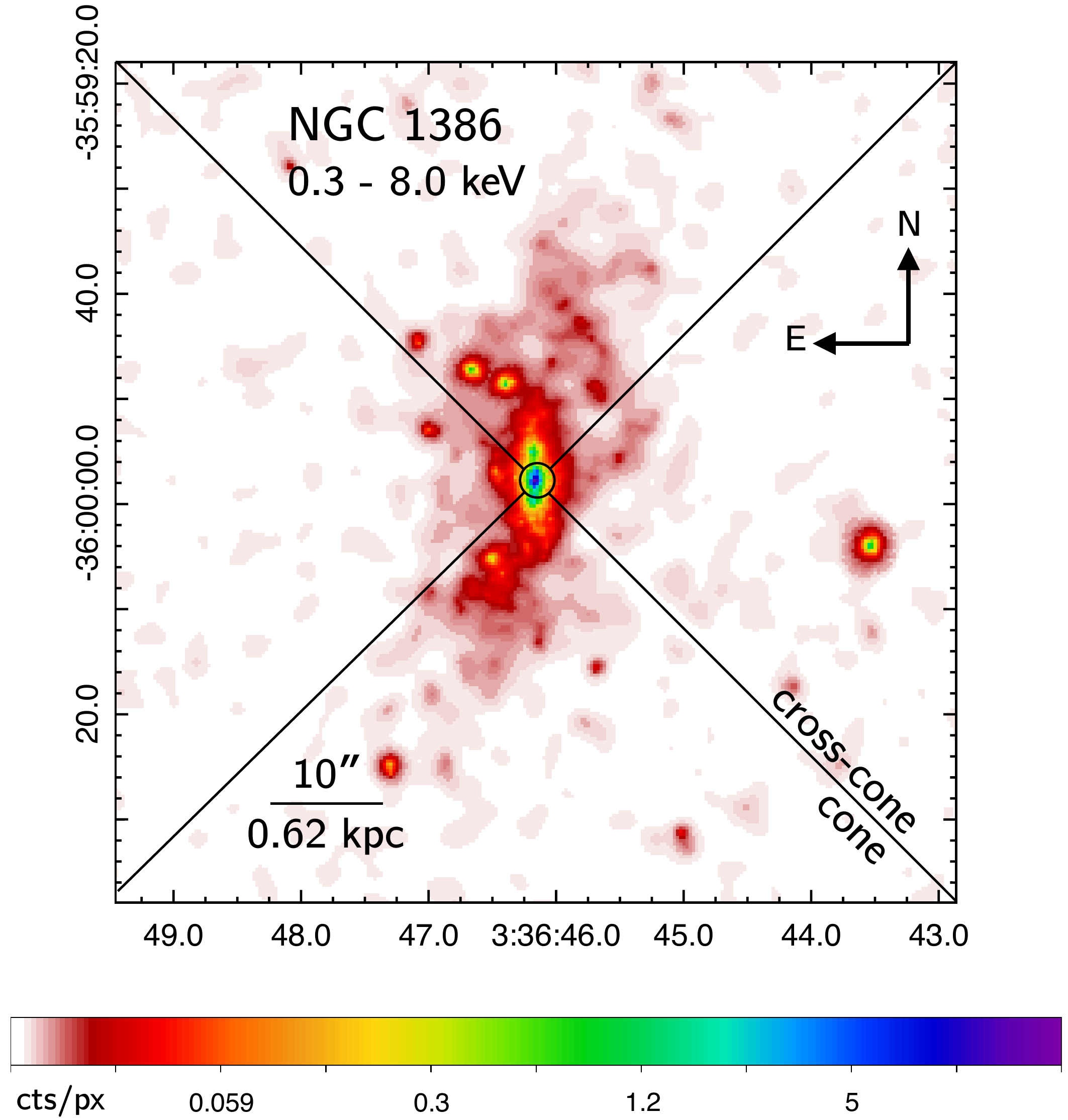}} \\
\caption{Merged $0.3-8.0$ keV \textit{Chandra} ACIS image of NGC 1386 with applied adaptive gaussian smoothing (\textit{dmingadapt}; $0.5-15$ pixel scales, 5 counts under kernel, 30 iterations) on image pixel $=1/8$ ACIS pixel. The image contours are logarithmic with colors corresponding to the number of counts per image pixel. The box size is 80\arcsec\, $\times$ 80\arcsec (4.96 $\times$ 4.96 kpc). Also shown are the 1.5\arcsec\,(0.093 kpc) circular region and cone/cross-cone quadrants used in our analysis of the X-ray extent. \label{fig:img:1386}}
\end{center}
\end{figure}
%%%%%%%%%%%%%%%%%%%%%

%%%%%%%%%%%%%%%%%%%%%
% TEXT
%%%%%%%%%%%%%%%%%%%%%
NGC 1386 is an SB0 type galaxy at RA=03:36:46.18 (54.192\degree), Dec=-35:59:57.87 (-35.999\degree), and $z=0.00290$ (D$_{\text{lum}}\sim$$12$ Mpc), with a water-megamaser (\citealt{Sch03mas}) and jet (\citealt{Nag99}). The AGN ($\log L_{x,2-10 keV}=41.84$ erg s$^{-1}$, \citealt{Bri15}) is optically classified as a Seyfert 2 (e.g., \citealt{Bri11}) with a Compton thick AGN ($N_H=5.61 \times 10^{24}$ cm$^{-2}$; \citealt{Bri15}). 
The full band ($0.3-8.0$ keV) adaptively smoothed image of NGC 1386 is shown in Figure \ref{fig:img:1386}. Previous observations of NGC 1386 have shown extended \ion{O}{3} narrow lines (e.g., \citealt{Sch03}) aligned along the north-south direction and coincident with our defined ``cone'' region as shown in Figure \ref{fig:img:1386}.

%%%%%%%%%%%%%%%%%%%%%%%%
\subsection{NGC 3393}
%%%%%%%%%%%%%%%%%%%%%%%%

%%%%%%%%%%%%%%%%%%%%%
% FIGURE : FULL IMAGE
\begin{figure}
\begin{center}
\resizebox{75mm}{!}{\includegraphics{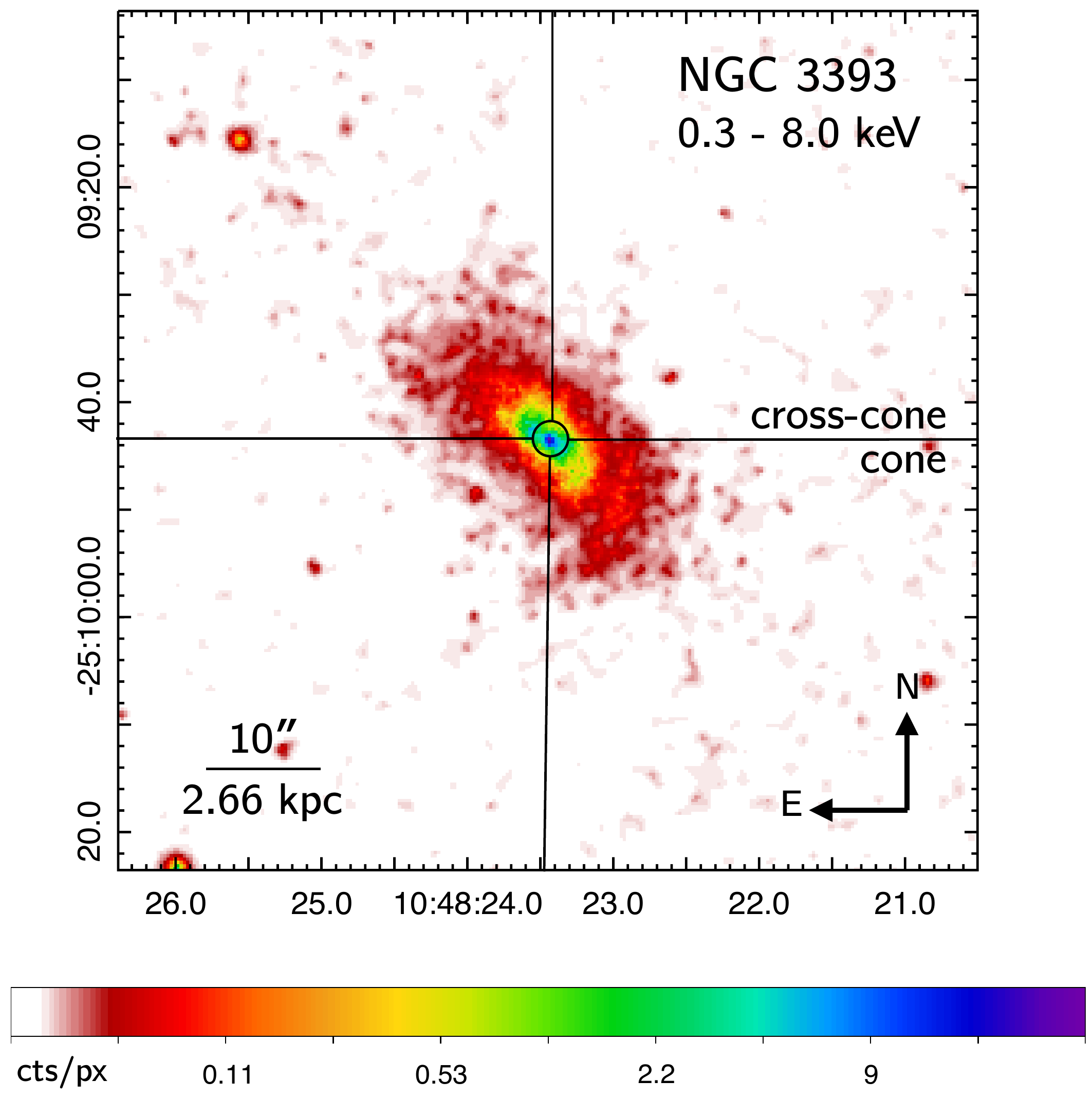}} \\
\caption{Merged $0.3-8.0$ keV \textit{Chandra} ACIS image of NGC 3393 with applied adaptive gaussian smoothing (\textit{dmingadapt}; $0.5-15$ pixel scales, 10 counts under kernel, 30 iterations) on image pixel $=1/8$ ACIS pixel. The image contours are logarithmic with colors corresponding to the number of counts per image pixel. The box size is 80\arcsec\, $\times$ 80\arcsec (21.28 $\times$ 21.28 kpc). Also shown are the 1.5\arcsec\,(0.399 kpc) circular region and cone/cross-cone quadrants used in our analysis of the X-ray extent. \label{fig:img:3393}}
\end{center}
\end{figure}
%%%%%%%%%%%%%%%%%%%%%

%%%%%%%%%%%%%%%%%%%%%
% TEXT
%%%%%%%%%%%%%%%%%%%%%
NGC 3393 is an SBab type galaxy at RA=10:48:23.46 (162.098\degree), Dec=-25:09:43.4 (-25.162\degree), and $z=0.01251$ (D$_{\text{lum}}\sim61$ Mpc), with a triple-lobed radio source and extended ionization cones oriented along the north-east direction (e.g., \citealt{Coo00}). The AGN is optically classified as a Seyfert 2 with a Compton thick obscuration ($N_H=1.897\times10^{24}$ cm$^{-2}$; \citealt{Mar18}; see also \citealt{Mai98,Gua05,Bur11,Kos15,Mak17}).
The full band ($0.3-8.0$ keV) adaptively smoothed image of NGC 3393 is shown in Figure \ref{fig:img:3393}. Consistent with previous X-ray observations (e.g., \citealt{Mak16,Mak17,Mak19}), we observe both the ``S'' shaped energetic central region and extended emission along P.A. 45\degree\, that we designate the ``cone region'', as shown in Figure \ref{fig:img:1386}.

%%%%%%%%%%%%%%%%%%%%%%%%
\subsection{NGC 5643}
%%%%%%%%%%%%%%%%%%%%%%%%

%%%%%%%%%%%%%%%%%%%%%
% FIGURE : FULL IMAGE
\begin{figure}
\begin{center}
\resizebox{75mm}{!}{\includegraphics{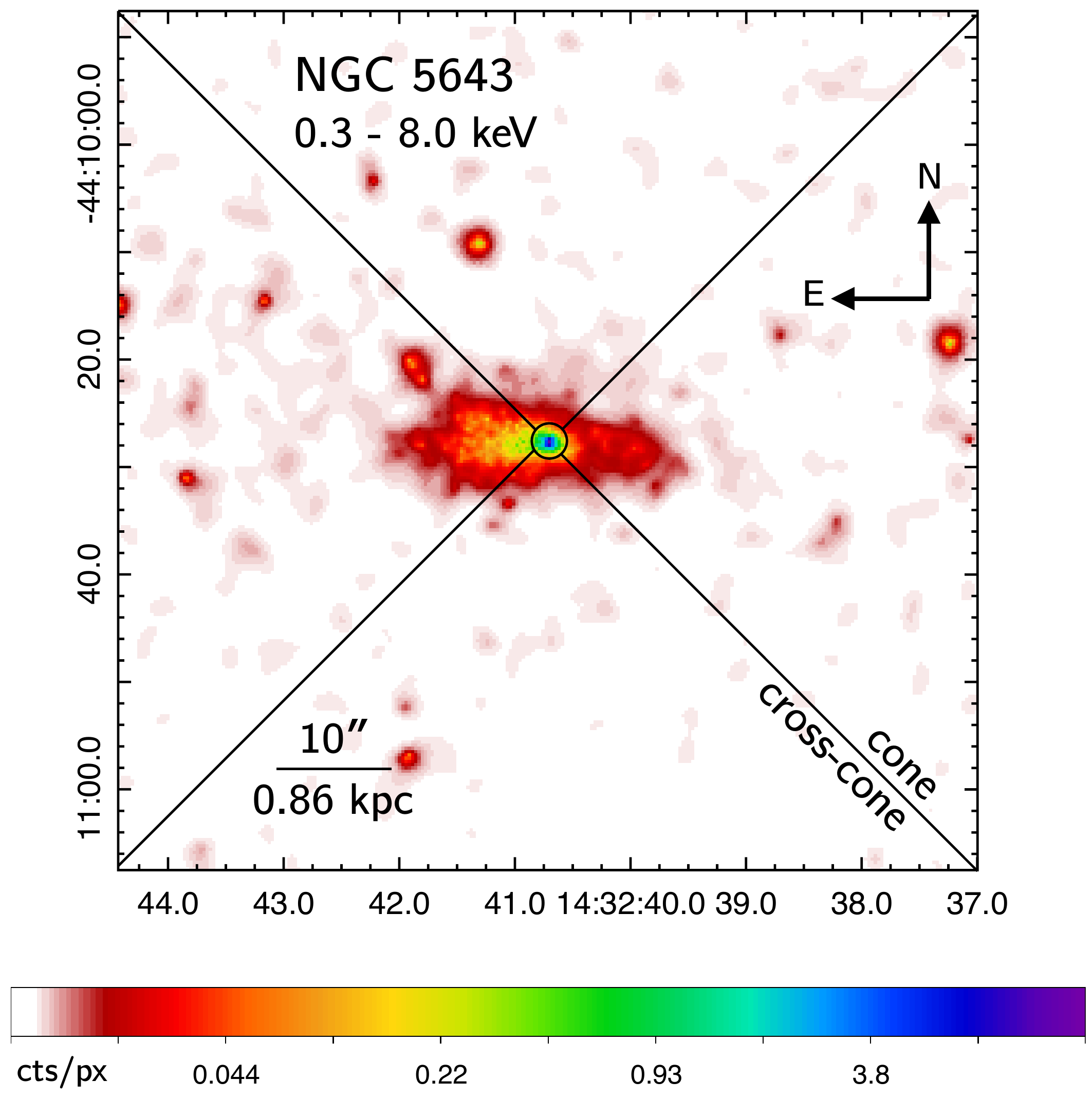}} \\
\caption{Merged $0.3-8.0$ keV \textit{Chandra} ACIS image of NGC 5643 with applied adaptive gaussian smoothing (\textit{dmingadapt}; $0.5-15$ pixel scales, 5 counts under kernel, 30 iterations) on image pixel $=1/8$ ACIS pixel. The image contours are logarithmic with colors corresponding to the number of counts per image pixel. The box size is 80\arcsec\, $\times$ 80\arcsec (6.88 $\times$ 6.88 kpc). Also shown are the 1.5\arcsec\,(0.129 kpc) circular region and cone/cross-cone quadrants used in our analysis of the X-ray extent. \label{fig:img:5643}}
\end{center}
\end{figure}
%%%%%%%%%%%%%%%%%%%%%

%%%%%%%%%%%%%%%%%%%%%
% TEXT
%%%%%%%%%%%%%%%%%%%%%
NGC 5643 is an SABc type galaxy at RA=14:32:40.74 (218.170\degree), Dec=-44:10:27.86 (-44.174\degree), and $z=0.00400$ (D$_{\text{lum}}\sim21$ Mpc), with a water maser (\citealt{Gre03}) and ultra-luminous X-ray source (e.g., \citealt{Ann15,Pin16}). The AGN ($L_{x,2-10 keV}=(0.8-1.7)\times10^{42}$ erg s$^{-1}$, \citealt{Ann15}) is optically classified as a Seyfert 2 (\citealt{Cid01}) with Compton thick obscuration ($N_H=1.594\times10^{24}$ cm$^{-2}$; \citealt{Mar18}; see also \citealt{Ris99,Gua04,Mai98,Bia06,Mat13,Ann15}) likely from a nuclear rotating obscuration disk (\citealt{Alo18}). 
The full band ($0.3-8.0$ keV) adaptively smoothed image of NGC 5643 is shown in Figure \ref{fig:img:5643}. We observe extended emission along the east-west direction, coincident with both the narrow line region ionization cone and kpc-scale radio lobes (\citealt{Mor85,Sch94,Fis13,Cre15}) which forms the basis for our cone/cross-cone regions (as indicated in Figure \ref{fig:img:5643}).

%%%%%%%%%%%%%%%%%%%%%%%%
\subsection{NGC 7212}
%%%%%%%%%%%%%%%%%%%%%%%%

%%%%%%%%%%%%%%%%%%%%%
% FIGURE : FULL IMAGE
\begin{figure}
\begin{center}
\resizebox{75mm}{!}{\includegraphics{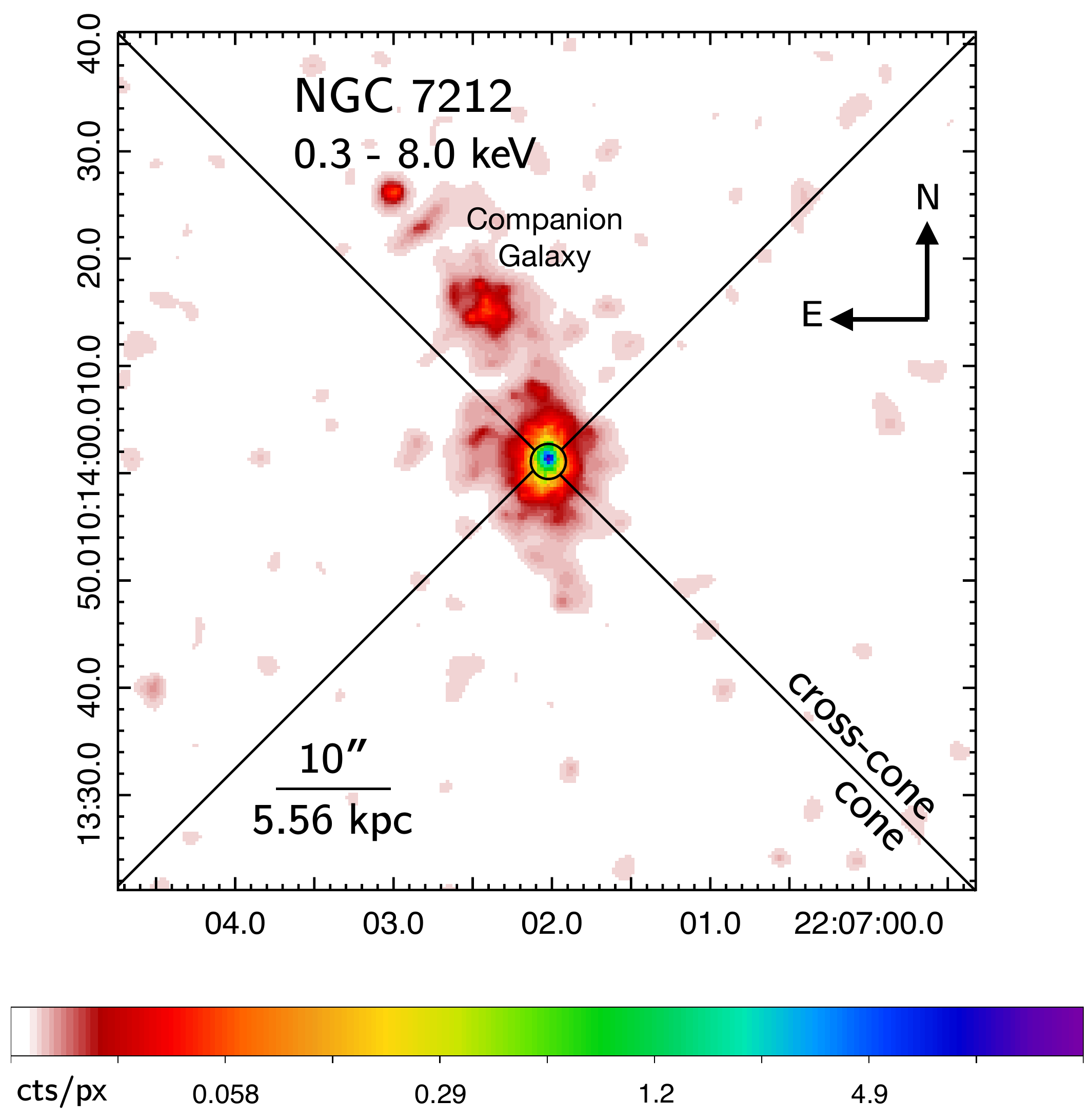}} \\
\caption{Merged $0.3-8.0$ keV \textit{Chandra} ACIS image of NGC 7212 with applied adaptive gaussian smoothing (\textit{dmingadapt}; $0.5-15$ pixel scales, 5 counts under kernel, 30 iterations) on image pixel $=1/8$ ACIS pixel. The image contours are logarithmic with colors corresponding to the number of counts per image pixel. The box size is 80\arcsec\, $\times$ 80\arcsec (44.48 $\times$ 44.48 kpc). Also shown are the 1.5\arcsec\,(0.834 kpc) circular region and cone/cross-cone quadrants used in our analysis of the X-ray extent. \label{fig:img:7212}}
\end{center}
\end{figure}
%%%%%%%%%%%%%%%%%%%%%

%%%%%%%%%%%%%%%%%%%%%
% TEXT
%%%%%%%%%%%%%%%%%%%%%
NGC 7212 is located at RA=22:07:01.30 (331.755\degree), Dec:+10:13:52 (10.231\degree), and $z=0.0266$ (D$_{\text{lum}}\sim115$ Mpc), in a compact group of three interacting galaxies (e.g., \citealt{Mun07}). The AGN ($\log M_{\text{bh}}=7.54$; $\log L/L_{Edd}=-1.55$; \citealt{Her15}) is optically classified as a Seyfert 2, with a kpc-scale extended narrow line region (ENLR; e.g., \citealt{Was81,Fal98,Sch03,Cra11,Con17}), and typical characteristics of Compton thick obscuration ($N_H=1.269\times10^{24}$ cm$^{-2}$; \citealt{Mar18}; see also \citealt{Ris00,Gua05,Lev06,Bia06,Sin11,Sev12,Her15}). 
The full band ($0.3-8.0$ keV) adaptively smoothed image of NGC 7212 is shown in Figure \ref{fig:img:7212}. The cone region indicated in Figure \ref{fig:img:7212} is coincident with the ENLR and aligned with the compact double radio source (extent $0.7\arcsec$; \citealt{Fal98,Dra03}).

%%%%%%%%%%%%%%%%%%%%%%%%%%%%%%%%%%%%%%%%%%%%%%%%%
% RADIAL PROFILES
%%%%%%%%%%%%%%%%%%%%%%%%%%%%%%%%%%%%%%%%%%%%%%%%%
\section{Spatial Analysis}\label{sec:rp}

%%%%%%%%%%%%%%%%%%%%%
% FIGURE : FWHM FITS
\begin{figure*}
\begin{center}
\begin{tabular}{c}
%\resizebox{80mm}{!}{\includegraphics{573_FWHM.pdf}} &  \resizebox{80mm}{!}{\includegraphics{1386_FWHM.pdf}} \\
%\resizebox{80mm}{!}{\includegraphics{3393_FWHM.pdf}} &  \resizebox{80mm}{!}{\includegraphics{5643_FWHM.pdf}} \\
%\multicolumn{2}{c}{\resizebox{80mm}{!}{\includegraphics{7212_FWHM.pdf}}} \\
\resizebox{160mm}{!}{\includegraphics{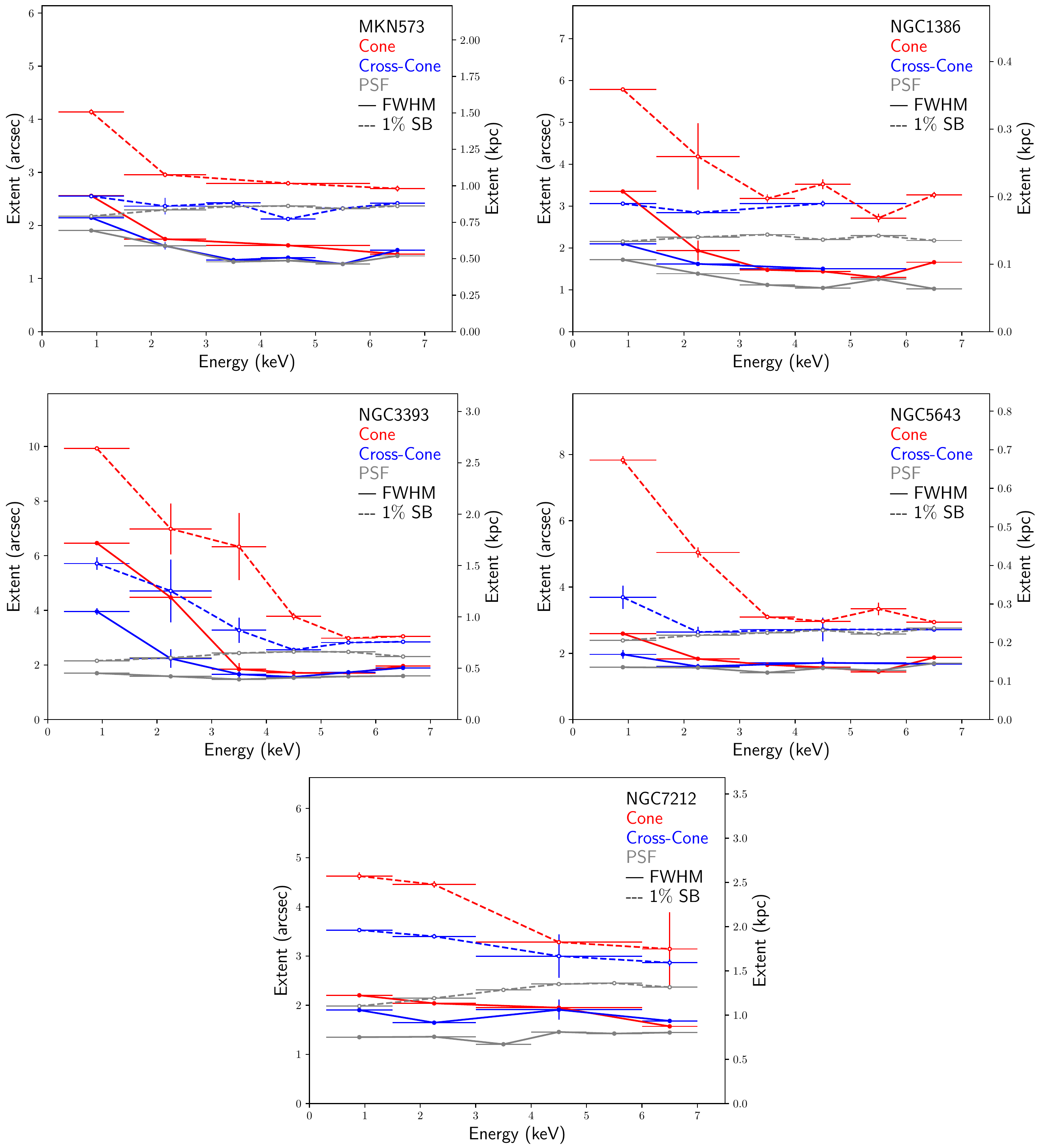}} \\
\end{tabular}
\caption{Emission extent as a function of energy calculated from the radial profiles for both the cone (red) and cross-cone (blue) regions of each CT AGN in our sample. The extent of the corresponding \textit{Chandra} PSF is shown in grey. Solid lines: Full width at half maximum surface brightness. Dashed lines: Full width at 1\% of the peak surface brightness. 1-sigma errors are shown. \label{fig:fwhm}}
\end{center}
\end{figure*}
%%%%%%%%%%%%%%%%%%%%%

%%%%%%%%%%%%%%%%%%%%%
% TEXT
%%%%%%%%%%%%%%%%%%%%%
The detailed morphologies that we are able to extract by capitalizing on the sub-pixel resolution of \textit{Chandra} enable a thorough investigation of the significance of the extended emission. Likewise, our sample is made of sources with $>2400$ counts, which allow us to break down the radial surface brightness profiles by cone and cross-cone regions to provide a better understanding of the origin of the emission and AGN geometry. As described in Section \ref{sec:obs}, we separate our images into four  90\degree quadrants (the biconical cone and cross-cone regions). Of the five sources, NGC 7212 is the only one to not exhibit a strong azimuthal dependence.

Using SAOImage ds9, we filtered our sources in eight energy bands and generated concentric annuli within each quadrant starting at the nucleus ($r=0.5$\arcsec) and working radially outward, increasing the width as necessary to maintain a minimum of 10 counts, until the regions became noise and background dominated (typically around 30\arcsec\, to 50\arcsec). The surface brightness profiles were extracted from these energy-and-quadrant-dependent regions (excluding obvious point sources) and background subtracted, before being compared to the \textit{Chandra} Point Source Functions (PSF) for the given energy band and regions (PSFs were created from an absorbed power-law spectrum typical for an AGN using ChaRT\footnote{http://cxc.harvard.edu/ciao/PSFs/chart2/} ($\Gamma\sim1.8$, $N_H$=$0.5\times10^{22}$ cm$^{-2}$) and MARX 5.4.0\footnote{https://space.mit.edu/cxc/marx/}, and following the CIAO simulation threads\footnote{http://cxc.harvard.edu/ciao/threads/psf.html}\footnote{http://cxc.harvard.edu/ciao/threads/marx\_sim/}). 

There are known uncertainties introduced when simulating the \textit{Chandra} PSF with ChaRT and MARX\footnote{https://cxc.harvard.edu/ciao/PSFs/chart2/caveats.html}, specifically, for energies $>2$ keV the wings of the simulated PSF underestimate the extent of the observed surface brightness profile. We estimate the potential uncertainty for two energy bands (2.4$-$2.6 keV and 6.4$-$6.6 keV) at two extents (10\arcsec\, and 100\arcsec) based on the analysis ``Wings of the \textit{Chandra} PSF''\footnote{https://cxc.harvard.edu/ccw/proceedings/02\_proc/presentations/t\_gaetz}. An extent of 100\arcsec\, is well outside of our area of interest that typically extends only to $\sim$30\arcsec$-\sim$50\arcsec\, before becoming noise dominated. 
For energies 2.4$-$2.6 keV, the factor difference between the simulated and observed surface brightness profiles at 10\arcsec\, is $\sim1.2$, and at 100\arcsec\, is $\sim1.7$. For energies 6.4$-$6.6 keV, the factor difference at 10\arcsec\, is $\sim1.9$ and at 100\arcsec is $\sim2.8$. Taking this extreme factor into account, we confirm that the 6.0$-$7.0 keV emission in all of our sources (excluding the cross-cone region of MKN 573) will continue to be significant ($>3\sigma$) in both the cone and cross-cone regions. However, we expect errors in the simulated PSF introduced by this method to be less than a factor of 2.8 at high energies and large extent. Additional discrepancies between the simulated and observed PSF associated with this method are mitigated using the recommended AspectBlur in MARX of 0.25\arcsec\, for ACIS-S observations. Despite these known uncertainties, we observe a significant difference in the source extent by cone angle, as described in the following sections, that cannot be solely attributed to poorly simulating the \textit{Chandra} PSF.

The \textit{Chandra} PSFs were energy filtered and normalized to the source counts in the nuclear ($<0.5$\arcsec) region before being subtracted from the source radial profiles to determine the quantity of total excess counts outside of the nuclear 0.5\arcsec\, region as a function of energy. We then calculate the ``total excess fraction'' by dividing the total excess counts by the total counts (not PSF subtracted) over the entire region of interest.
Similarly, we quantify an extended excess by adding up the PSF subtracted counts outside of a 1.5\arcsec\, region as a function of energy. By increasing the radius to 1.5\arcsec\, from 0.5\arcsec\,, we are further limiting any potential contamination from the CT AGN. We then calculate the ``extended fraction'' by dividing the extended excess counts by the total counts (not PSF subtracted) over the entire region of interest.

To better compare the extent in each energy band, we then calculate the full width at half maximum (FWHM) and the full width at 1$\%$ of the surface brightness (in log space) for each radial profile (following, e.g., \citealt{Fab18II,Jon20}). This not only normalizes the brightness of each energy band, it also minimizes the bias between each source in our sample of five CT AGN caused by variations in signal-to-noise. These width calculations are made by fitting the radial profiles with a spline approximation (or for profiles with less than four points, a gaussian curve) with errors derived from a bootstrap Monte Carlo analysis.

%%%%%%%%%%%%%%%%%%%%%%%%
\subsection{MKN 573}
%%%%%%%%%%%%%%%%%%%%%%%%

MKN 573 exhibits extended emission with an azimuthal dependence along the designated ``cone'' region that transitions into a circular blob at the higher energies (Figure \ref{fig:grid:573}). The surface brightness in the cone and cross-cone lies predominately above the \textit{Chandra} PSF, and exhibits bumps and waves outside of $\sim10$\arcsec\, that may correspond to point sources that were unaccounted for in the radial profile extraction. 
For $0.3-8.0$ keV, we find $3483\pm59$ counts at $>0.5$\arcsec\, above the \textit{Chandra} PSF in the cone region, and $978\pm31$ in cross-cone region (Table \ref{tab:cts:573}). Between $6-7$ keV, where we would expect to see Fe K fluorescence, we find significant counts above the PSF in both the cone and cross-cone region ($29\pm5$ counts and $10\pm3$ counts, respectively).

The total excess fraction for MKN 573 in $0.3-8.0$ keV is $55.3\pm1.2$ \% in the cone region, and $24.2\pm0.9$ \% in the cross-cone region, which is consistent with the observed azimuthal dependence. In the hard X-rays between $6-7$ keV, the difference in the excess fraction is more stark at $29.9\pm6.3$ \% in the cone and $9.0\pm3.0$ \% in the cross-cone regions.
More interesting, however, is the total extended fraction, as it probes the excess emission farther from the influence of the CT AGN. We find an extended fraction for the $0.3-8.0$ keV to be $40.6\pm1.0$ \% in the cone and $13.1\pm0.6$ \% in the cross-cone regions. Probing the hard X-rays at $6-7$ keV, we find an extended fraction of $16.1\pm4.4$ \% in the cone and $1.8\pm1.3$ \% in cross-cone regions. These three metrics for the X-ray extent are all significant in the cone region and cross-cone region, with the exception of the extended fraction in the cross-cone region.

Further exploring the excess as a function of energy, we calculated the FWHM for each radial profile at each energy bin (Figure \ref{fig:fwhm}; top, left). We find that the cone and cross cone are extended in the soft X-rays $\sim0.4$ kpc and $\sim0.25$ kpc farther than in the harder X-rays, respectively, although the slope of this extent as a function of energy is fairly shallow, especially above $\sim3$ keV. 
At 1\% of the surface brightness (Figure \ref{fig:fwhm}; top, left), where we better probe the extended emission, we find that the extent of the cross-cone region does not significantly change as a function of energy. The cone region, however, is extended $\sim0.5$ kpc farther in the soft energies compared to the hard energies.

%%%%%%%%%%%%%%%%%%%%%
% TABLE : COUNTS
\begin{table*}
\centering
\caption{\textbf{MKN 573 $-$} Excess counts over the \textit{Chandra} PSF (normalized to the central 0.5\arcsec) for select energy bands.}\label{tab:cts:573}
\begin{scriptsize}
\begin{tabular}{ccccccccccc}
\hline\hline
Energy & \multicolumn{2}{c}{Total Counts} & \multicolumn{2}{c}{Counts $>0.5$\arcsec} & \multicolumn{2}{c}{Counts $>1.5$\arcsec} & \multicolumn{2}{c}{Extended Fraction} & \multicolumn{2}{c}{Total Excess Fraction} \\
(kev) & Cone & Cross-cone & Cone & Cross-cone & Cone & Cross-cone & Cone & Cross-cone & Cone & Cross-cone \\
\hline
0.3-1.5 & $5377\pm73$ & $3138\pm56$ & $3399\pm56$ & $774\pm28$ & $2312\pm48$ & $404\pm20$ & $43.0\pm1.1$ & $12.9\pm0.7$ & $57.6\pm1.3$ & $24.7\pm1.0$ \\
1.5-3.0 & $591\pm24$ & $477\pm22$ & $261\pm16$ & $81\pm9$ & $197\pm14$ & $69\pm8$ & $33.2\pm2.7$ & $14.5\pm1.9$ & $44.2\pm3.3$ & $16.9\pm2.0$ \\
3.0-4.0 & $115\pm11$ & $82\pm9$ & $39\pm6$ & $21\pm5$ & $27\pm5$ & $16\pm4$ & $23.4\pm5.0$ & $19.6\pm5.3$ & $33.4\pm6.2$ & $25.8\pm6.3$ \\
4.0-5.0 & $69\pm8$ & $52\pm7$ & $16\pm4$ & $3\pm2$ & $4\pm2$ & $0\pm0$ & $5.9\pm3.0$ & $0.0\pm0.0$ & $23.6\pm6.5$ & $5.5\pm3.3$ \\
5.0-6.0 & $55\pm7$ & $66\pm8$ & $13\pm4$ & $15\pm4$ & $9\pm3$ & $14\pm4$ & $16.4\pm5.9$ & $21.7\pm6.3$ & $22.9\pm7.2$ & $22.7\pm6.5$ \\
6.0-7.0 & $97\pm10$ & $107\pm10$ & $29\pm5$ & $10\pm3$ & $16\pm4$ & $2\pm1$ & $16.1\pm4.4$ & $1.8\pm1.3$ & $29.9\pm6.3$ & $9.0\pm3.0$ \\
7.0-8.0 & $10\pm3$ & $15\pm4$ & $3\pm2$ & $8\pm3$ & $3\pm2$ & $8\pm3$ & $28.9\pm19.8$ & $53.9\pm23.9$ & $28.9\pm19.8$ & $53.9\pm23.9$ \\
\hline
0.3-8.0 & $6304\pm79$ & $4037\pm64$ & $3483\pm59$ & $978\pm31$ & $2562\pm51$ & $529\pm23$ & $40.6\pm1.0$ & $13.1\pm0.6$ & $55.3\pm1.2$ & $24.2\pm0.9$ \\
\hline
\end{tabular}
\end{scriptsize}
\end{table*}
%%%%%%%%%%%%%%%%%%%%%

%%%%%%%%%%%%%%%%%%%%%%%%
\subsection{NGC 1386}
%%%%%%%%%%%%%%%%%%%%%%%%

Compared to MKN 573, NGC 1386 presents a more challenging picture due to additional point sources primarily aligned with the cone region, even at the harder X-rays (Figure \ref{fig:grid:1386}). Despite this, we successfully extracted the radial profiles in each energy band (Figure \ref{fig:rp:1386}) and find, unsurprisingly, that the central region is dominated by the point source. Outside of $\sim7-10$\arcsec\, however, the surface brightness falls at a gentler slope.

We calculate the excess counts above the PSF at $>0.5$\arcsec\, and find, for $0.3-8.0$ keV, $3352\pm58$ in the cone and $886\pm30$ excess counts in the cross-cone region (Table \ref{tab:cts:1386}). Compared to the total counts in this energy band, we find a total excess fraction of $71.8\pm1.6$ \% in the cone and $48.0\pm2.0$ \% in the cross-cone.
In the hard X-rays between $6-7$ keV, we find $82\pm9$ ($\sigma$) in the cone and $34\pm6$ in the cross-cone region. This corresponds to a total excess fraction of $54.9\pm7.6$ \% in the cone and $89.6\pm21.1$ \% in the cross-cone.
For the extended emission above 1.5\arcsec, we find a total extended fraction of $49.6\pm1.3$ \% in the cone and $28.3\pm1.4$ \% in the cross-cone region for $0.3-8.0$ keV. In $6-7$ keV, the total extended fraction is $17.2\pm3.7$ \% in the cone and $27.2\pm9.5$ \% in the cross-cone. Similar to MKN 573, we find significant excess emission across all three metrics, with the exception of the cross-cone extended fraction which is just shy of a $3\sigma$ result.

The radial profile FWHM as a function of energy for NGC 1386 is similar to what was found for MKN 573 (Figure \ref{fig:fwhm}; top, right). The cone and cross cone are more extended at soft energies than hard energies with a difference of $0.1$ kpc and $\sim0.04$ kpc, respectively, but remain relatively flat above $\sim2$ keV. At 1\% the surface brightness we also find that the cross-cone region exhibits a consistent extent across the energy bands. For the cone region, the soft X-rays are extended $\sim0.15$ kpc farther than both the cone hard energies and the cross-cone region.

%%%%%%%%%%%%%%%%%%%%%
% TABLE : COUNTS
\begin{table*}
\centering
\caption{\textbf{NGC 1386 $-$} Excess counts over the \textit{Chandra} PSF (normalized to the central 0.5\arcsec) for select energy bands.}\label{tab:cts:1386}
\begin{scriptsize}
\begin{tabular}{ccccccccccc}
\hline\hline
Energy & \multicolumn{2}{c}{Total Counts} & \multicolumn{2}{c}{Counts $>0.5$\arcsec} & \multicolumn{2}{c}{Counts $>1.5$\arcsec} & \multicolumn{2}{c}{Extended Fraction} & \multicolumn{2}{c}{Total Excess Fraction} \\
(kev) & Cone & Cross-cone & Cone & Cross-cone & Cone & Cross-cone & Cone & Cross-cone & Cone & Cross-cone \\ \hline
0.3-1.5 & $3747\pm61$ & $1520\pm39$ & $2783\pm53$ & $713\pm27$ & $2022\pm45$ & $438\pm21$ & $53.9\pm1.5$ & $28.8\pm1.6$ & $74.3\pm1.9$ & $46.9\pm2.1$ \\
1.5-3.0 & $480\pm22$ & $200\pm14$ & \multicolumn{1}{l}{$298\pm17$} & \multicolumn{1}{l}{$93\pm10$} & $172\pm13$ & $58\pm8$ & $35.9\pm3.2$ & $29.1\pm4.3$ & $62.0\pm4.6$ & $46.4\pm5.8$ \\
3.0-4.0 & $132\pm12$ & $40\pm6$ & $89\pm9$ & $19\pm4$ & $60\pm8$ & $15\pm4$ & $45.3\pm7.1$ & $37.1\pm11.3$ & $67.5\pm9.3$ & $47.4\pm13.3$ \\
4.0-5.0 & $83\pm9$ & $18\pm4$ & $51\pm7$ & $13\pm4$ & $28\pm5$ & $0\pm0$ & $33.3\pm7.3$ & $0.0\pm0.0$ & $62.0\pm11.0$ & $71.1\pm26.3$ \\
5.0-6.0 & $63\pm8$ & $21\pm5$ & $32\pm6$ & $4\pm2$ & $15\pm4$ & $2\pm1$ & $23.3\pm6.7$ & $8.7\pm6.7$ & $51.0\pm11.1$ & $19.9\pm10.6$ \\
6.0-7.0 & $148\pm12$ & $38\pm6$ & $82\pm9$ & $34\pm6$ & $26\pm5$ & $10\pm3$ & $17.2\pm3.7$ & $27.2\pm9.5$ & $54.9\pm7.6$ & $89.6\pm21.1$ \\
7.0-8.0 & $17\pm4$ & $9\pm3$ & $17\pm4$ & $10\pm3$ & $5\pm2$ & $8\pm3$ & $30.9\pm15.3$ & $97.0\pm35.1$ & $98.6\pm33.6$ & $119.7\pm55.2$ \\ \hline
0.3-8.0 & $4667\pm68$ & $1846\pm43$ & $3352\pm58$ & $886\pm30$ & $2313\pm48$ & $523\pm23$ & $49.6\pm1.3$ & $28.3\pm1.4$ & $71.8\pm1.6$ & $48.0\pm2.0$ \\ \hline
\end{tabular}
\end{scriptsize}
\end{table*}
%%%%%%%%%%%%%%%%%%%%%

%%%%%%%%%%%%%%%%%%%%%%%%
\subsection{NGC 3393}
%%%%%%%%%%%%%%%%%%%%%%%%

NGC 3393 is elongated in the ``cone'' region (Figure \ref{fig:grid:3393}) and continues to have an azimuthal dependence even in the hard X-rays, although to a lesser extent. The radial profiles (Figure \ref{fig:grid:3393}) that we extract are peaky nearest the central point source before exhibiting a bump outside of $\sim10$\arcsec\, in the cone region, similar to MKN 573 and NGC 1386.

From these radial profiles, we calculate the excess counts above the \textit{Chandra} PSF and find the most counts out of our entire sample, $13080\pm114$ counts in the cone and $3785\pm62$ in the cross cone for $0.3-8.0$ keV (Table \ref{tab:cts:3393}). Similarly, in the hard X-rays between $6-7$ keV, we find $203\pm14$ excess counts in the cone and $169\pm13$ counts in the cross-cone.
The total excess fraction for these energy bands are $88.2\pm1.1$ \% in the cone and $70.4\pm1.5$ \% in the cross-cone for $0.3-8.0$ keV, and $30.8\pm2.5$ \% in the cone and $30.5\pm2.7$ \% in the cross-cone for $6-7$ keV. Likewise, the extended fractions are $70.1\pm0.9$ \% in the cone and $41.9\pm1.1$ \% in the cross-cone, and $12.4\pm1.5$ \% in the cone and $9.2\pm1.3$ \% in the cross-cone region. All three metrics are significant in the cone and cross-cone regions. This is further evidence of extended hard X-ray emission in NGC 3393, as first suggested in \citet{Mak17} and later in \citet{Mak19}, and consistent with wind-driven shocks and collisionally excited gas.

The FWHM extent deviates from that of MKN 573 and NGC 1386 (Figure \ref{fig:fwhm}; middle, left). Below 3 keV, there is a much more significant slope in the measured extent. The cone extends $>1$ kpc in the soft X-rays than the hard X-rays, and $>0.5$ kpc compared to the cross-cone in the same soft energy bin. Similarly, there is a steep slope at 1\% of the surface brightness, but that flattens at higher energies than the FWHM. The cone region is extended $>1$ kpc more than the cross-cone at the softest energies and is $\sim1.9$ kpc more extended than the hard band.

%%%%%%%%%%%%%%%%%%%%%
% TABLE : COUNTS
\begin{table*}
\centering
\caption{\textbf{NGC 3393 $-$} Excess counts over the \textit{Chandra} PSF (normalized to the central 0.5\arcsec) for select energy bands.}\label{tab:cts:3393}
\begin{scriptsize}
\begin{tabular}{ccccccccccc}
\hline\hline
Energy & \multicolumn{2}{c}{Total Counts} & \multicolumn{2}{c}{Counts $>0.5$\arcsec} & \multicolumn{2}{c}{Counts $>1.5$\arcsec} & \multicolumn{2}{c}{Extended Fraction} & \multicolumn{2}{c}{Total Excess Fraction} \\
(kev) & Cone & Cross-cone & Cone & Cross-cone & Cone & Cross-cone & Cone & Cross-cone & Cone & Cross-cone \\ \hline
0.3-1.5 & $10019\pm100$ & $2839\pm53$ & $9550\pm98$ & $2531\pm50$ & $7764\pm88$ & $1554\pm39$ & $77.5\pm1.2$ & $54.7\pm1.7$ & $95.3\pm1.4$ & $89.1\pm2.4$ \\
1.5-3.0 & $2805\pm53$ & $1047\pm32$ & \multicolumn{1}{l}{$2434\pm49$} & \multicolumn{1}{l}{$781\pm28$} & $1952\pm44$ & $530\pm23$ & $69.6\pm2.1$ & $50.6\pm2.7$ & \multicolumn{1}{l}{$86.8\pm2.4$} & \multicolumn{1}{l}{$74.6\pm3.5$} \\
3.0-4.0 & $464\pm22$ & $261\pm16$ & $375\pm19$ & $154\pm12$ & $297\pm17$ & $117\pm11$ & $64.0\pm4.8$ & $44.9\pm5.0$ & $80.9\pm5.6$ & $59.1\pm6.0$ \\
4.0-5.0 & $328\pm18$ & $244\pm16$ & $166\pm13$ & $58\pm8$ & $117\pm11$ & $45\pm7$ & $35.8\pm3.9$ & $18.4\pm3.0$ & $50.7\pm4.8$ & $23.8\pm3.5$ \\
5.0-6.0 & $301\pm17$ & $281\pm17$ & $102\pm10$ & $82\pm9$ & $59\pm8$ & $48\pm7$ & $19.5\pm2.8$ & $17.0\pm2.7$ & $34.0\pm3.9$ & $29.2\pm3.7$ \\
6.0-7.0 & $661\pm26$ & $555\pm24$ & $203\pm14$ & $169\pm13$ & $82\pm9$ & $51\pm7$ & $12.4\pm1.5$ & $9.2\pm1.3$ & $30.8\pm2.5$ & $30.5\pm2.7$ \\
7.0-8.0 & $53\pm7$ & $51\pm7$ & $46\pm7$ & $30\pm6$ & $35\pm6$ & $19\pm4$ & $66.8\pm14.6$ & $37.5\pm10.1$ & $87.2\pm17.6$ & $60.2\pm13.8$ \\ \hline
0.3-8.0 & $14835\pm122$ & $5375\pm73$ & $13080\pm114$ & $3785\pm62$ & $10400\pm102$ & $2254\pm48$ & $70.1\pm0.9$ & $41.9\pm1.1$ & $88.2\pm1.1$ & $70.4\pm1.5$ \\ \hline
\end{tabular}
\end{scriptsize}
\end{table*}
%%%%%%%%%%%%%%%%%%%%%

%%%%%%%%%%%%%%%%%%%%%%%%
\subsection{NGC 5643}
%%%%%%%%%%%%%%%%%%%%%%%%

NGC 5643 has an azimuthal dependence, as in the other CT AGN discussed thus far, however, this dependence is not symmetric around the central point source (Figure \ref{fig:grid:5643}). There exists an interesting soft X-ray excess in the East-cone region. As the energy increases, this asymmetry lessens and even transitions to a slight excess in the West-cone region in the hard energy bands. The radial profiles are consistent with this picture and show lopsided curves surrounding the central point source (Figure \ref{fig:rp:5643}).

The total excess counts in the $0.3-8.0$ keV band are $2880\pm54$ in the cone region and $1052\pm32$ in the cross cone region (Table \ref{tab:cts:5643}). At the $6-7$ keV energies, we find $93\pm10$ counts in the cone $38\pm6$ counts in cross-cone region. Based on these counts, the total excess fraction is $75.3\pm1.9$ \% in the cone and $53.9\pm2.1$ \% in the cross-cone for $0.3-8.0$ keV, and $38.1\pm4.6$ \% in the cone and $14.6\pm2.6$ \% in the cross-cone region for $6-7$ keV. Meanwhile, the total extended fractions are $64.0\pm1.7$ \% in the cone and $43.8\pm1.8$ \% in the cross-cone region for $0.3-8.0$ keV, and $15.8\pm2.7$ \% in the cone and $8.0\pm1.8$ \% in the cross-cone region for $6-7$ keV. All three metrics are significant in the cone and cross-cone regions.

We find that the calculated FWHM for both the cone and cross-cone region are consistent with a flat relationship with energy, other than at the softest energy band where they differ by $<1$ kpc (Figure \ref{fig:fwhm}; middle, right). At 1\% of the surface brightness, the cone region is extended $\sim0.4$ kpc farther in the soft than hard X-rays, while the cross cone is similar to FWHM, not changing above 1.5 keV. Furthermore, the cone region extends $\sim0.38$ kpc farther than the cross-cone at 1\% of the surface brightness.

%%%%%%%%%%%%%%%%%%%%%
% TABLE : COUNTS
\begin{table*}
\caption{\textbf{NGC 5643 $-$} Excess counts over the \textit{Chandra} PSF (normalized to the central 0.5\arcsec) for select energy bands.}\label{tab:cts:5643}
\begin{scriptsize}
\centering
\begin{tabular}{ccccccccccc}
\hline\hline
Energy & \multicolumn{2}{c}{Total Counts} & \multicolumn{2}{c}{$>0.5$\arcsec} & \multicolumn{2}{c}{$>1.5$\arcsec} & \multicolumn{2}{c}{Extended Fraction} & \multicolumn{2}{c}{Total Excess Fraction} \\
(kev) & Cone & Cross-cone & Cone & Cross-cone & Cone & Cross-cone & Cone & Cross-cone & Cone & Cross-cone \\ \hline
0.3-1.5 & $2288\pm48$ & $812\pm29$ & $2013\pm45$ & $588\pm24$ & $1762\pm42$ & $477\pm22$ & $77.0\pm2.4$ & $58.7\pm3.4$ & $88.0\pm2.7$ & $72.3\pm3.9$ \\
1.5-3.0 & $653\pm26$ & $376\pm19$ & \multicolumn{1}{l}{$481\pm22$} & \multicolumn{1}{l}{$198\pm14$} & $424\pm21$ & $190\pm14$ & $64.9\pm4.0$ & $50.6\pm4.5$ & \multicolumn{1}{l}{$73.6\pm4.4$} & \multicolumn{1}{l}{$52.6\pm4.6$} \\
3.0-4.0 & $220\pm15$ & $147\pm12$ & $126\pm11$ & $70\pm8$ & $94\pm10$ & $61\pm8$ & $42.5\pm5.2$ & $41.3\pm6.3$ & $57.0\pm6.4$ & $47.6\pm6.9$ \\
4.0-5.0 & $202\pm14$ & $197\pm14$ & $109\pm10$ & $76\pm9$ & $81\pm9$ & $76\pm9$ & $40.3\pm5.3$ & $38.3\pm5.2$ & $53.9\pm6.4$ & $38.8\pm5.2$ \\
5.0-6.0 & $152\pm12$ & $110\pm11$ & $71\pm8$ & $29\pm5$ & $34\pm6$ & $29\pm5$ & $22.5\pm4.3$ & $15.9\pm4.1$ & $47.0\pm6.7$ & $26.3\pm5.5$ \\
6.0-7.0 & $244\pm16$ & $256\pm16$ & $93\pm10$ & $38\pm6$ & $39\pm6$ & $21\pm5$ & $15.8\pm2.7$ & $8.0\pm1.8$ & $38.1\pm4.6$ & $14.6\pm2.6$ \\
7.0-8.0 & $61\pm8$ & $46\pm7$ & $47\pm7$ & $33\pm6$ & $37\pm6$ & $27\pm5$ & $61.3\pm12.7$ & $58.2\pm14.1$ & $77.3\pm15.0$ & $72.0\pm16.4$ \\ \hline
0.3-8.0 & $3826\pm62$ & $1950\pm44$ & $2880\pm54$ & $1052\pm32$ & $2448\pm50$ & $856\pm29$ & $64.0\pm1.7$ & $43.9\pm1.8$ & $75.3\pm1.9$ & $53.9\pm2.1$ \\ \hline
\end{tabular}
\end{scriptsize}
\end{table*}
%%%%%%%%%%%%%%%%%%%%%

%%%%%%%%%%%%%%%%%%%%%%%%
\subsection{NGC 7212}
%%%%%%%%%%%%%%%%%%%%%%%%

Unlike the other CT AGN in this sample, NGC 7212 does not exhibit a strong azimuthal dependence (Figure \ref{fig:grid:7212}), thus we select the ``cone'' region to align with the optically classified extended narrow line emission region (e.g., \citealt{Con17}). Extracting the radial profiles was made more challenging due to contamination from the companion galaxies in the North-cone region and the soft X-ray filaments that connect them. They exhibit a similar ``bump'' to the other CT AGN in the inner $\sim$\arcsec, but due to its distance, these potential wings are not well resolved (Figure \ref{fig:rp:7212}).

The excess counts above the PSF in $0.3-8.0$ keV are  $1359\pm37$ in the cone and $731\pm27$ in the cross cone (Table \ref{tab:cts:7212}). This corresponds to a total excess fraction of $54.9\pm1.9$ \% in the cone and $41.6\pm1.8$ \% in the cross-cone region. Focusing on the extended emission, the total extended fraction is $30.3\pm1.3$ \% in the cone and $18.8\pm1.1$ \% in the cross-cone region. At the hard energy band, $6-7$ keV, we find counts in excess of $54\pm7$ in the cone and $42\pm7$ in the cross-cone region. The excess fraction for this hard band is $27.4\pm4.2$ \% in the cone and $23.3\pm4.0$ \% in the cross-cone. 
Comparing the excess fraction to the extended fraction, we find for $6-7$ keV, a total extended fraction of $8.8\pm2.2$ \% in the cone and $5.6\pm1.8$ \% in the cross-cone. All three metrics are significant in the cone and cross-cone regions.

For NGC 7212, the FWHM in arcsec as a function of energy appears to be very much consistent with the other CT AGN in this sample, although the extent in kpc is the largest in the sample (Figure \ref{fig:fwhm}; bottom). For the cone region, we find an extent of $\sim1.5$ kpc in the softest band and a $\sim0.45$ kpc difference in the extent between the soft and the hard X-rays. The cross-cone region, however, is much more consistent. At 1\% of the surface brightness, the extent is flatter for the cross-cone than the cone region, especially above 3 keV. The cone is $\sim0.9$ kpc more extended in the soft X-rays than both the corresponding cross-cone soft band and the cone hard band. 

%%%%%%%%%%%%%%%%%%%%%
% TABLE : COUNTS
\begin{table*}
\centering
\caption{\textbf{NGC 7212 $-$} Excess counts over the \textit{Chandra} PSF (normalized to the central 0.5\arcsec) for select energy bands.}\label{tab:cts:7212}
\begin{scriptsize}
\begin{tabular}{ccccccccccc}
\hline\hline
Energy & \multicolumn{2}{c}{Total Counts} & \multicolumn{2}{c}{$>0.5$\arcsec} & \multicolumn{2}{c}{$>1.5$\arcsec} & \multicolumn{2}{c}{Extended Fraction} & \multicolumn{2}{c}{Total Excess Fraction} \\
(kev) & Cone & Cross-cone & Cone & Cross-cone & Cone & Cross-cone & Cone & Cross-cone & Cone & Cross-cone \\ \hline
0.3-1.5 & $868\pm30$ & $631\pm25$ & $552\pm24$ & $328\pm18$ & $339\pm18$ & $172\pm13$ & $39.0\pm2.5$ & $27.3\pm2.3$ & $63.7\pm3.5$ & $52.0\pm3.5$ \\
1.5-3.0 & $688\pm26$ & $429\pm21$ & \multicolumn{1}{l}{$393\pm20$} & \multicolumn{1}{l}{$141\pm12$} & $221\pm15$ & $75\pm9$ & $32.1\pm2.5$ & $17.5\pm2.2$ & \multicolumn{1}{l}{$57.1\pm3.6$} & \multicolumn{1}{l}{$32.9\pm3.2$} \\
3.0-4.0 & $252\pm16$ & $181\pm13$ & $119\pm11$ & $80\pm9$ & $56\pm8$ & $24\pm5$ & $22.2\pm3.3$ & $13.0\pm2.8$ & $47.4\pm5.3$ & $44.1\pm5.9$ \\
4.0-5.0 & $221\pm15$ & $153\pm12$ & $101\pm10$ & $35\pm6$ & $48\pm7$ & $17\pm4$ & $21.7\pm3.5$ & $11.1\pm2.8$ & $45.8\pm5.5$ & $22.8\pm4.3$ \\
5.0-6.0 & $207\pm14$ & $147\pm12$ & $73\pm9$ & $47\pm7$ & $42\pm7$ & $14\pm4$ & $20.4\pm3.4$ & $9.8\pm2.7$ & $35.4\pm4.8$ & $31.9\pm5.3$ \\
6.0-7.0 & $196\pm14$ & $180\pm13$ & $54\pm7$ & $42\pm7$ & $17\pm4$ & $10\pm3$ & $8.8\pm2.2$ & $5.6\pm1.8$ & $27.4\pm4.2$ & $23.3\pm4.0$ \\
7.0-8.0 & $43\pm7$ & $39\pm6$ & $18\pm4$ & $27\pm5$ & $13\pm4$ & $22\pm5$ & $30.1\pm9.6$ & $55.7\pm14.8$ & $40.9\pm11.6$ & $67.9\pm17.0$ \\ \hline
0.3-8.0 & $2475\pm50$ & $1760\pm42$ & $1359\pm37$ & $731\pm27$ & $751\pm27$ & $331\pm18$ & $30.3\pm1.3$ & $18.8\pm1.1$ & $54.9\pm1.9$ & $41.6\pm1.8$ \\ \hline
\end{tabular}
\end{scriptsize}
\end{table*}
%%%%%%%%%%%%%%%%%%%%%

%%%%%%%%%%%%%%%%%%%%%%%%%%%%%%%%%%%%%%%%%%%%%%%%%
% DISCUSSON
%%%%%%%%%%%%%%%%%%%%%%%%%%%%%%%%%%%%%%%%%%%%%%%%%
\section{Discussion}\label{sec:dis}

%%%%%%%%%%%%%%%%%%%%%
% TABLE:
\begin{table*}[]
\centering
\caption{Total combined excess fraction and observed properties of our CT AGN.}\label{tab:prop}
\begin{tabular}{cccccccc}
\hline\hline
         & Total Excess Fraction & log $N_H$\tablenotemark{(a)}                        & d$_{25}$\tablenotemark{(b)}       & log $M_{BH}$\tablenotemark{(c)} & log $L_{x,2-10 keV}$\tablenotemark{(d)} & log $\nu L_{\nu,12\mu m}$\tablenotemark{(e)} & $\lambda_{Edd}$\tablenotemark{(f)}\\
Source   & Cone+Cross-cone  & (cm$^{-2}$)                        & (kpc)          & (M$_\odot$)   & (erg s$^{-1}$)    & (erg s$^{-1}$)              \\ \hline
MKN 573  & $43.1\pm0.8$     & $>24.2$                          & $28.33\pm0.06$ & $7.37$       & $41.54$          & $43.53\pm0.08$ 	& 0.0105 \\
NGC 1386 & $65.1\pm1.3$     & $24.70\pm0.09$                   & $5.29\pm0.004$ & $7.24$       & $41.60$          & $42.39\pm0.09$            & 0.0010\\
NGC 3393 & $83.4\pm0.9$     & $24.28\substack{+0.09 \\ -0.04}$ & $21.73\pm0.02$ & $7.48$       & $41.29$          & $42.88\pm0.08$            & 0.0018\\
NGC 5643 & $68.1\pm1.4$     & $24.20\substack{+0.11 \\ -0.08}$ & $7.56\pm0.004$ & $6.30$       & $40.87$          & $42.52\pm0.12$            & 0.0121\\
NGC 7212 & $49.3\pm1.3$     & $24.10\substack{+0.11 \\ -0.08}$ & $42.78\pm0.09$ & $7.54$       & $42.60$          & $43.65\pm0.15$    & 0.0092\\ \hline       
\end{tabular}
\begin{flushleft}\tablecomments{\tablenotemark{(a)} \citealt{Gua05,Bri15,Mar18}; \tablenotemark{(b)} \citealt{deV91}; \tablenotemark{(c)} \citealt{Bia07,Kon08,Her15}; \tablenotemark{(d)} \citealt{Gon15,Her15}; \tablenotemark{(e)} \citealt{Asm14}; \tablenotemark{(f)} $\lambda_{Edd}=\nu L_{\nu,12\mu m}/L_{Edd}$, where $L_{Edd}=1.38\times10^{38}*M_{BH}$ and $\nu L_{\nu,12\mu m}$ is used as a proxy for $L_{bol}$.} \end{flushleft}
\end{table*}
%%%%%%%%%%%%%%%%%%%%%

%%%%%%%%%%%%%%%%%%%%%
% FIGURE : 1% SURFACE BRIGHTNESS
\begin{figure*}
\begin{center}
\begin{tabular}{cc}
\resizebox{160mm}{!}{\includegraphics{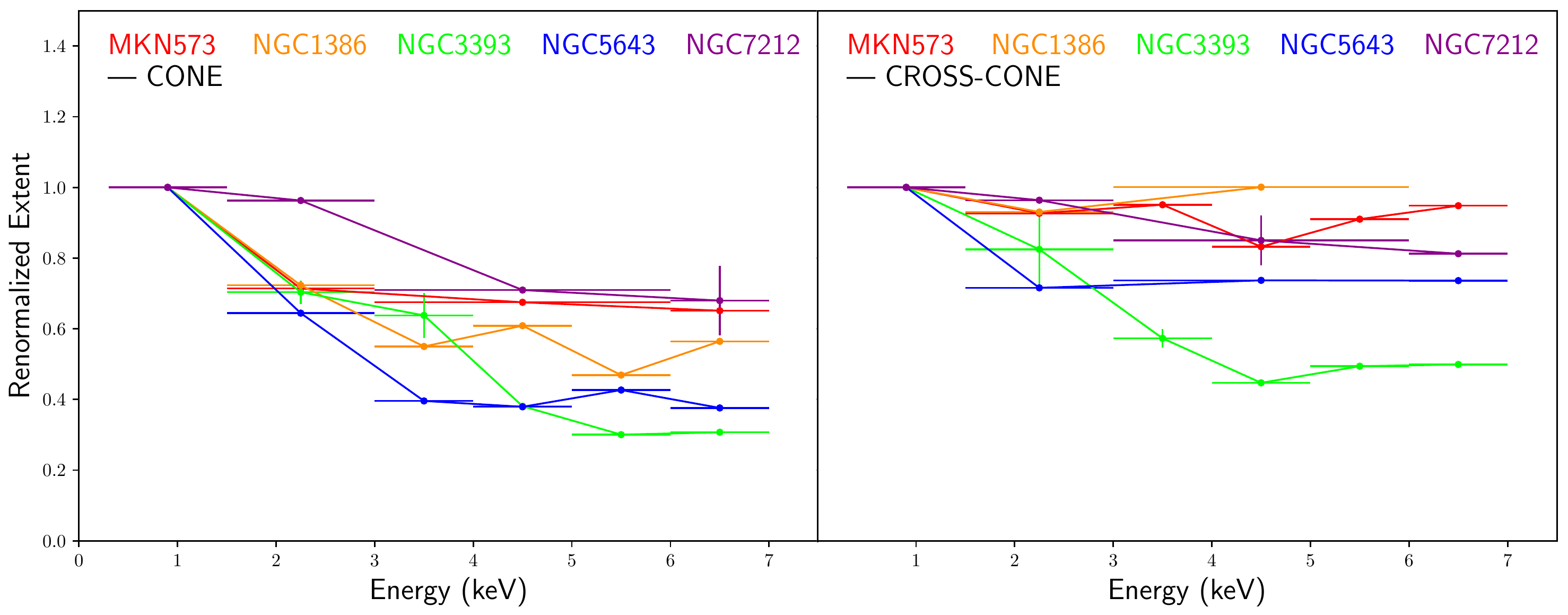}}
\end{tabular}
\caption{Renormalized extent (kpc) at 1\% of the surface brightness for our sample of five CT AGN as a function of energy: (left) cone regions (right) cross-cone regions. \label{fig:sb}}
\end{center}
\end{figure*}
%%%%%%%%%%%%%%%%%%%%%

%%%%%%%%%%%%%%%%%%%%%
% FIGURE : SLOPE PROPERTIES
\begin{figure*}
\begin{center}
\begin{tabular}{cc}
\resizebox{110mm}{!}{\includegraphics{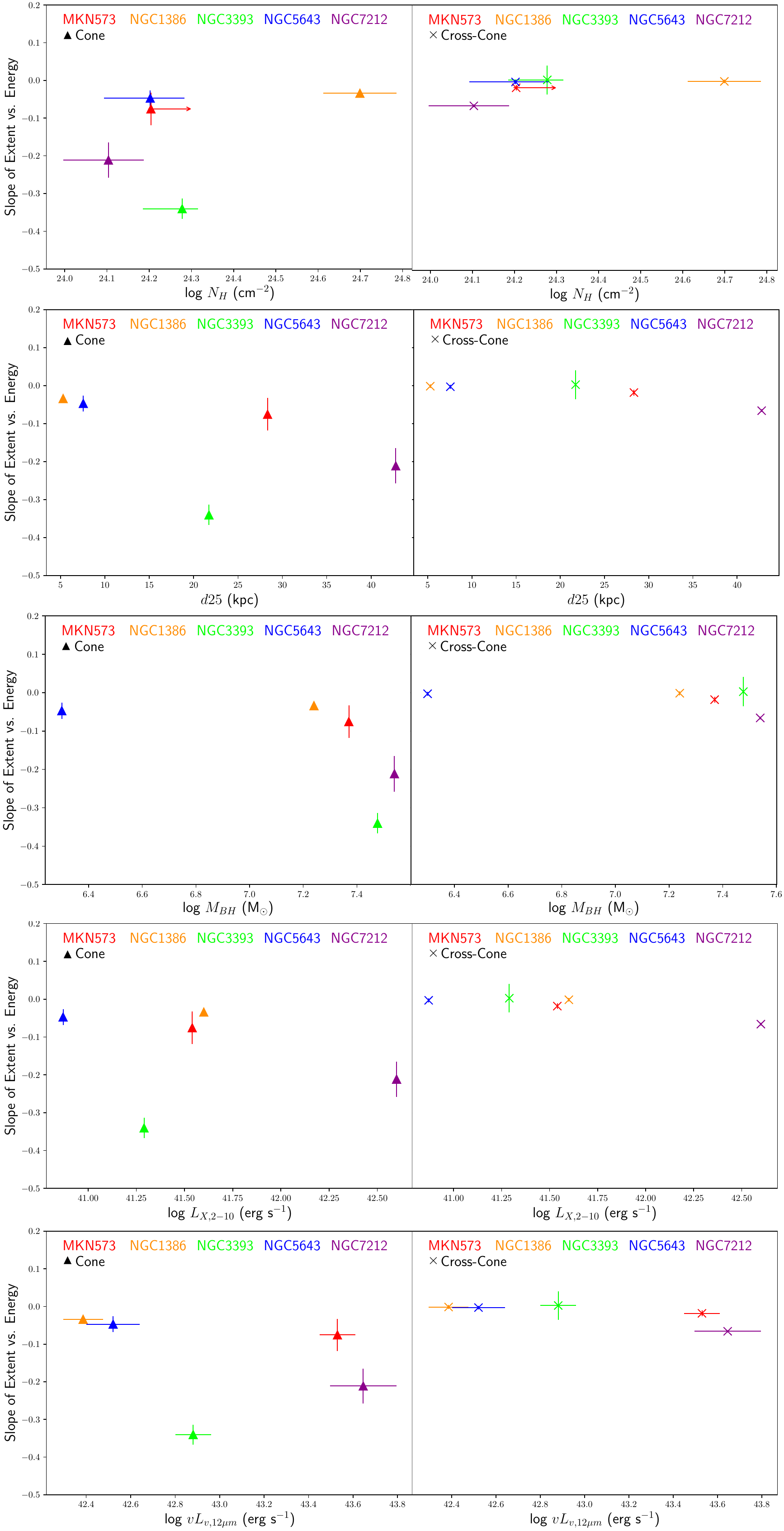}} \\
\end{tabular}
\caption{Best-fit slope of the extent at 1\% of the surface brightness versus energy in the cone and cross-cone regions (Figure \ref{fig:sb}) for each of the five CT AGN as a function of observed properties: (top to bottom) $\log N_H$, d$_{25}$, $\log M_{BH}$, $\log L_{x,2-10 keV}$, $\log \nu L_{\nu,12 \mu m}$. For each characteristic property, we find that the extent in the cross-cone region is less likely to be influenced by energy (i.e., the slope is consistent with 0). We observe a potentially interesting trend with black hole mass in the cone region (center, left). \label{fig:slope}}
\end{center}
\end{figure*}
%%%%%%%%%%%%%%%%%%%%%

\subsection{Spatial Extent at 1\% Surface Brightness}

As discussed in Section \ref{sec:rp}, we use the radial profile width at 1\% of the surface brightness as a proxy measurement for the X-ray extent of our CT AGN. 
For the cone region, we find that the CT AGN with the largest measured extent is NGC 3393 with $extent_{1\%}=2.6$ kpc in the soft X-rays. The next closest is NGC 7212, the most distant source in our sample, with $extent_{1\%}=2.55$ kpc in the soft X-rays. The remaining sources peak at $1.5$ kpc (MKN 574), $0.67$ kpc (NGC 5643), and $0.36$ kpc (NGC 1386).
In the cross-cone region, NGC 7212 is the most extended at $1.9$ kpc. This is less surprising since NGC 7212 exhibits the least azimuthally dependent morphology. The other sources peak at $1.5$ kpc (NGC 3393), $0.9$ (MKN 573), $0.3$ kpc (NGC 5643), and $0.19$ kpc (NGC 1386).

To better compare the slopes of these extent-energy relationships, we renormalized the extent of each source in kpc to the softest X-ray band (Figure \ref{fig:sb}), and fit the slopes with a simple line model. We find that the cone regions exhibit steeper drops in the X-ray extent with increasing energy, averaging a $\sim50$ \% drop by $6-7$ keV. 
This energy dependence is similar to that found in ESO 428-G014 (\citealt{Fab18II}), and suggests that the molecular clouds scattering photons are consistent across the CT AGN population and richer at smaller galactic radii (this is also observed in Milky Way molecular clouds; e.g., \citealt{Nak06}).
In comparison, the cross-cone region is much flatter (with the exception of NGC 3393 that more closely matches the cone regime). Excluding NGC 3393, the cross-cone extent drops by $\sim10$ \%, on average, by $6-7$ keV. 

The stronger dependence on energy in the cone region compared to the cross-cone may be explained by an inclination effect of the ionization cone with respect to the galaxy disk. 
The ionization cone itself often arises as AGN photons propagate through and interact with the ISM in the galaxy disk (e.g., \citealt{Sch96}). Soft X-rays are emitted via the photoionization of the diffuse, gaseous ISM and through collisional ionization in the presence of a jet (e.g., MKN 573, \citealt{Pag12}), while the hard X-rays originate from the interaction of AGN photons with dense molecular clouds (e.g., \citealt{Rey97}). Based on our findings where the soft X-rays extend farther than the hard X-rays, we can surmise that there is a galactocentric radial dependence on the size and density of the disk molecular clouds, such that the largest, densest clouds where the hard X-rays originate are closest to the nucleus.

In this case, one can imagine that if there is an ionization cone fully aligned with and propagating through the disk, one would observe the energy dependence we see in the cone region. There would be no dependence in the cross cone region since the AGN photons do not interact with the disk ISM, and are solely interacting with the flattened disk of the innermost dense molecular clouds. 
Likewise, if the ionization cone propagates perpendicular to the host disk, there would not be an energy dependence in the cone. It would be possible, however, for the cross-cone to exhibit some energy dependence, although the emission would be suppressed by attenuation from the torus, compared to the cone region in the first example.
That said, in the case of NGC 3393, where we observe an energy dependence in the cone and cross-cone region, it is possible that the ionization cone (and by definition, cross-cone) has some characteristic inclination  with respect to the galaxy disk such that the AGN photons from both regions interact with the inner dense molecular clouds, as well as the disk ISM at larger radii.

We further analyzed the slopes of the extent as a function of observed host galaxy and black hole properties; obscuring column density, host galaxy diameter (d$_{25}$), black hole mass, X-ray luminosity ($2-10$ keV), and 12$\mu m$ Luminosity (Table \ref{tab:prop}; Figure \ref{fig:slope}). For each characteristic property, we find little to no slope in the cross-cone region. In the cone region, the slopes are more diverse. 
Interestingly, the slope for NGC 3393 is very steep, deviating from a possible trend made by the other four sources in $N_H$, d$_{25}$, $\log L_{x,2-10}$, and $\log \nu L_{\nu,12\mu m}$. For all five sources, however, there may exist an interesting correlation between $M_{BH}$ and the slope of the extent as a function of energy, such that for higher black hole masses, the soft X-ray extent dominates the hard X-rays.

\subsection{Spatial Extent as a Function of Observables}

Since our sample is made up of ``nearby'' CT AGN, the availability of multiwavelength data provides a unique advantage to probe the X-ray extent as a function of AGN and host galaxy properties (Table \ref{tab:prop}). We split our sample into their respective cone and cross-cone regions and then calculate the total excess fraction for these regions in the wide-band $0.3-8.0$ keV and the hard-band $3.0-7.0$ keV. When applicable, we compare these results with those from \citet{Ma20}, although we note that the excess fractions are calculated only to 8$\arcsec\,$ in $0.3-7.0$ keV and $3.0-7.0$ keV. Furthermore, only two of the \citet{Ma20} sources (NGC 3281 and ESO 137-G034) had enough counts to be separated into a cone and cross-cone region, while the remaining sources are treated in entirety.

\subsubsection{Column Density}

%%%%%%%%%%%%%%%%%%%%%
% FIGURE : FEX PROPERTIES : NH
\begin{figure*}
\begin{center}
\begin{tabular}{cc}
\resizebox{115mm}{!}{\includegraphics{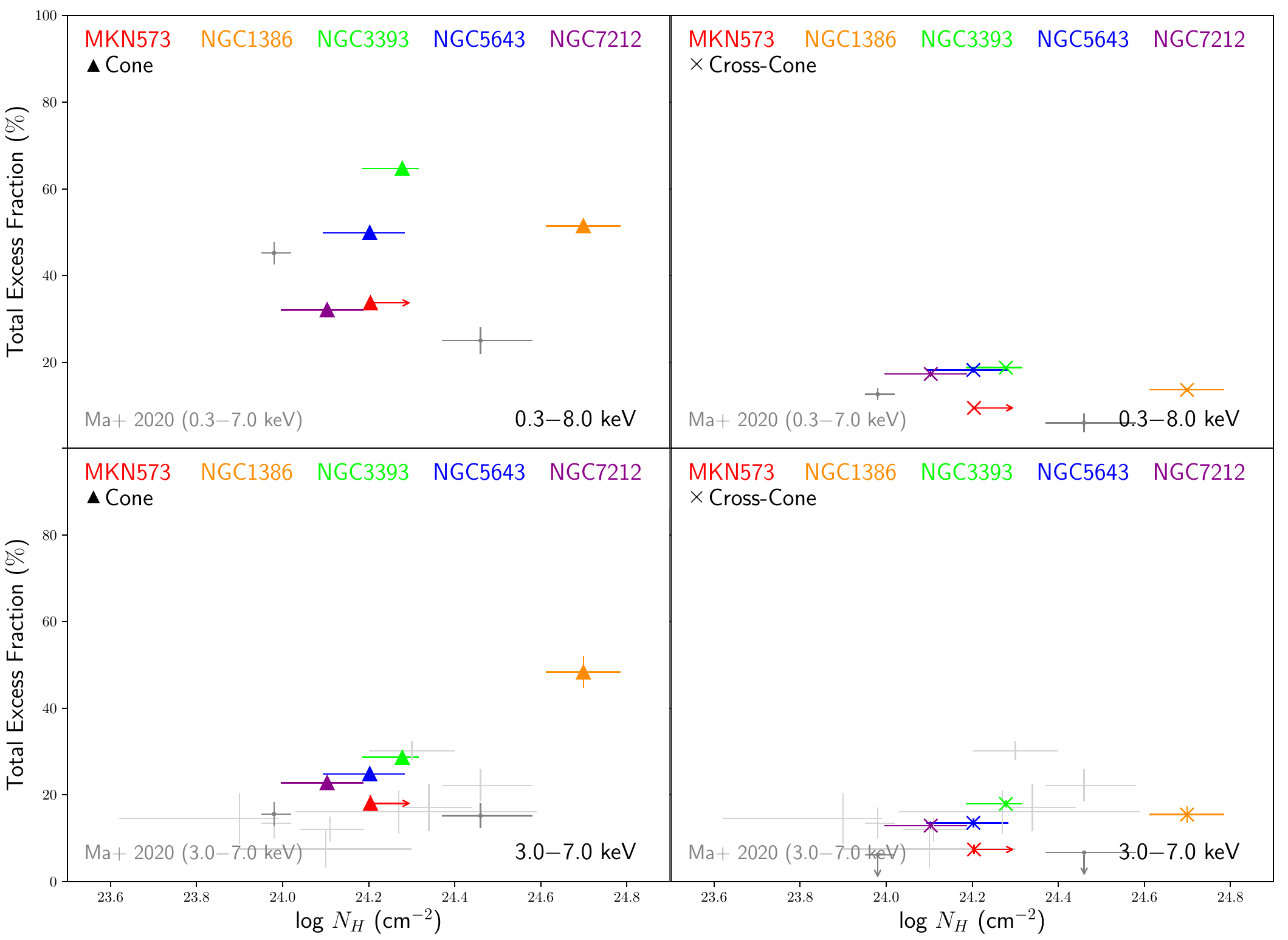}} \\
\end{tabular}
\caption{Total excess fraction in the cone (left) and cross-cone (right) regions for each of the five CT AGN as a function of column density ($\log N_H$; Table \ref{tab:prop}) for both the wide-band ($0.3-8.0$ keV; top) and hard-band ($3.0-7.0$ keV; bottom). We include the sources from \citet{Ma20} for comparison with the $3.0-7.0$ keV, but note that the excess fraction is calculated for the cone and cross-cone regions for only two sources (grey), while we include eight sources for which the excess fraction is calculated for the entire circular region (light grey). We find that the excess fraction in the cone region is more dispersed in the wide-band than the hard-band and exhibits a slight trend in the hard-band consistent with \citet{Ma20}. The cross-cone region, however, is fairly flat across both energies. \label{fig:fex:nh}}
\end{center}
\end{figure*}
%%%%%%%%%%%%%%%%%%%%%

We first probe the dependence of the X-ray extent on the obscuring column density (Figure \ref{fig:fex:nh}). We calculate the Pearson coefficient\footnote{https://docs.scipy.org/doc/scipy/reference/generated/scipy.stats.pearsonr.html} of this dependence in each region for each energy band, noting that our sources only cover a limited range in column density clustered around $\log N_H \sim 24$ and more sources with diverse column densities are required to better probe this correlation. In $0.3-8.0$ keV we find a Pearson coefficient of 0.42 (0.09 including the \citealt{Ma20} sources) in the cone region, i.e., weakly and not significantly correlated, respectively. In $3.0-7.0$ keV, however, we find a strong correlation in the cone region (coefficient of 0.95; 0.71 including the \citealt{Ma20} sources) between $N_H$ and the total excess hard X-ray fraction that is stronger than that found with the \citet{Ma20} sources alone (coefficient of 0.42), although the strength of this correlation is mostly driven by NGC 1386. 
In the cross-cone region for both the wider $0.3-8.0$ keV band, and $3.0-7.0$ keV hard X-ray band, we find only a slight correlation with coefficients of -0.24 (0.38 including the \citealt{Ma20} sources) and -0.26 (0.34 including the \citealt{Ma20} sources), respectively.
This suggests that the nuclear obscuration may be correlated with the abundance of molecular clouds in the disk. The cross-cone region, however, does not exhibit this trend. This may be explained by a torus with obscuration that dominates in this plane over any effects from molecular clouds, or more simply that molecular clouds are not coupled with obscuration of the torus in the disk plane.

\subsubsection{Host Galaxy Diameter}

%%%%%%%%%%%%%%%%%%%%%
% FIGURE : FEX PROPERTIES : d25
\begin{figure*}
\begin{center}
\begin{tabular}{cc}
\resizebox{115mm}{!}{\includegraphics{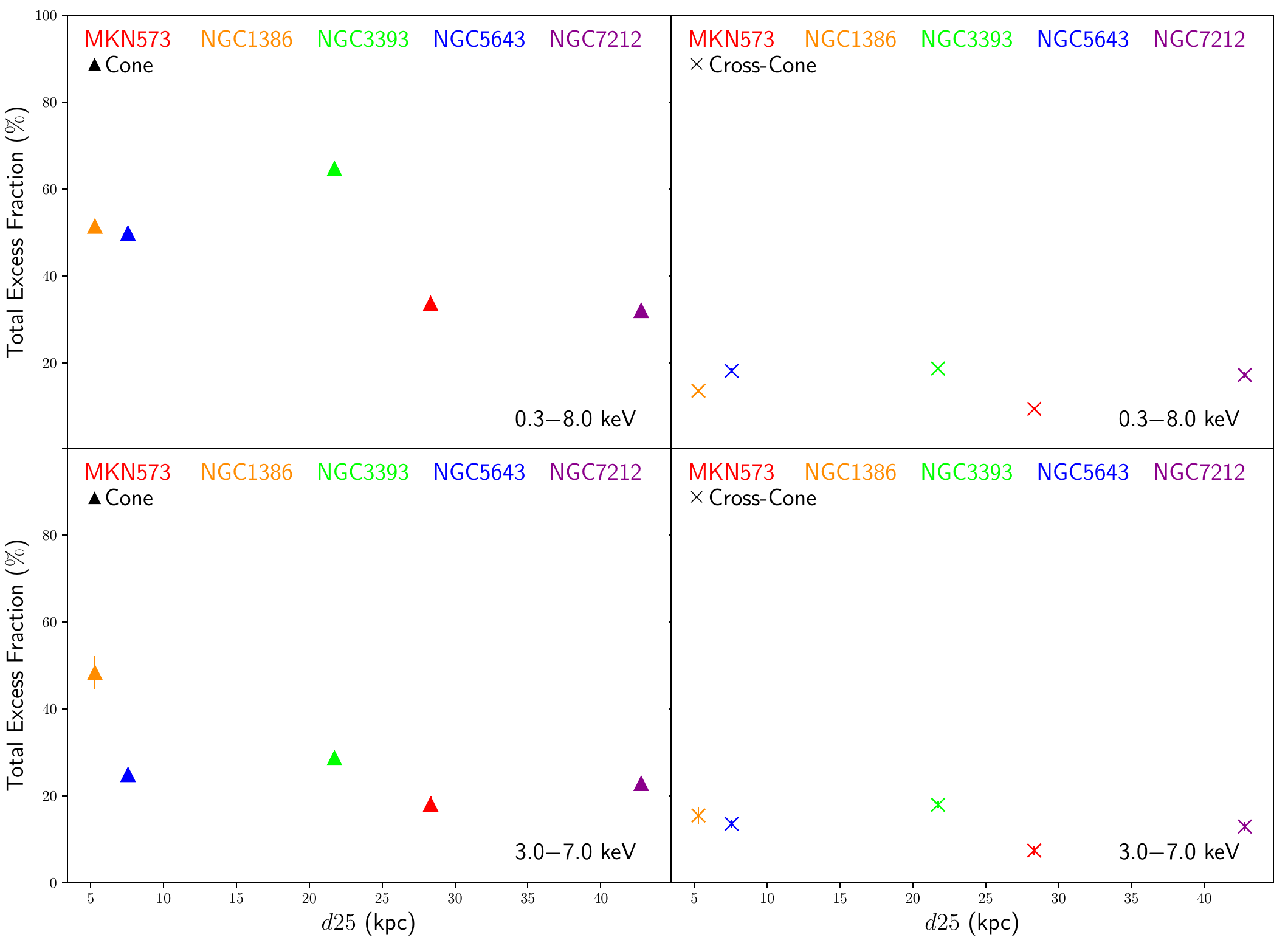}} \\
\end{tabular}
\caption{Total excess fraction in the cone (left) and cross-cone (right) regions for each of the five CT AGN as a function of host galaxy diameter (d$_{25}$; Table \ref{tab:prop}) for both the wide-band ($0.3-8.0$ keV; top) and hard-band ($3.0-7.0$ keV; bottom). We find some indication of a negative trend in excess fraction with increasing d$_{25}$ for the cone region, but none in the cross-cone region.
\label{fig:fex:d25}}
\end{center}
\end{figure*}
%%%%%%%%%%%%%%%%%%%%%

For each source, we compare the excess fraction with the d$_{25}$ diameter measure of the host galaxy from the \textit{Third Reference Catalog of Bright Galaxies} (\citealt{deV91}), as shown in Figure \ref{fig:fex:d25}. 
We calculate the Pearson coefficient of this dependence in each region for each energy band.
In both the full-band and hard-band cone region, we find the excess fraction shallowly decreases with increasing d$_{25}$ (Pearson coefficients of -0.62 and -0.64, respectively). This may imply that the extent of the galaxy is uncoupled from the extent of the AGN emission, such that as the host galaxy increases in size, there is not any additional energy from the central source imparted into the host galaxy to maintain a constant excess fraction. This may be further complicated by evolutionary effects, such that in the presence of strong outflows, the bicone may be more spatially developed.
As with $N_H$, there is no significant trend observed in the cross-cone region (Pearson coefficients of -0.04 and -0.36 in $0.3-8.0$ keV and $3.0-7.0$ keV, respectively).

\subsubsection{Black Hole Mass}

%%%%%%%%%%%%%%%%%%%%%
% FIGURE : FEX PROPERTIES : MBH
\begin{figure*}
\begin{center}
\begin{tabular}{cc}
\resizebox{115mm}{!}{\includegraphics{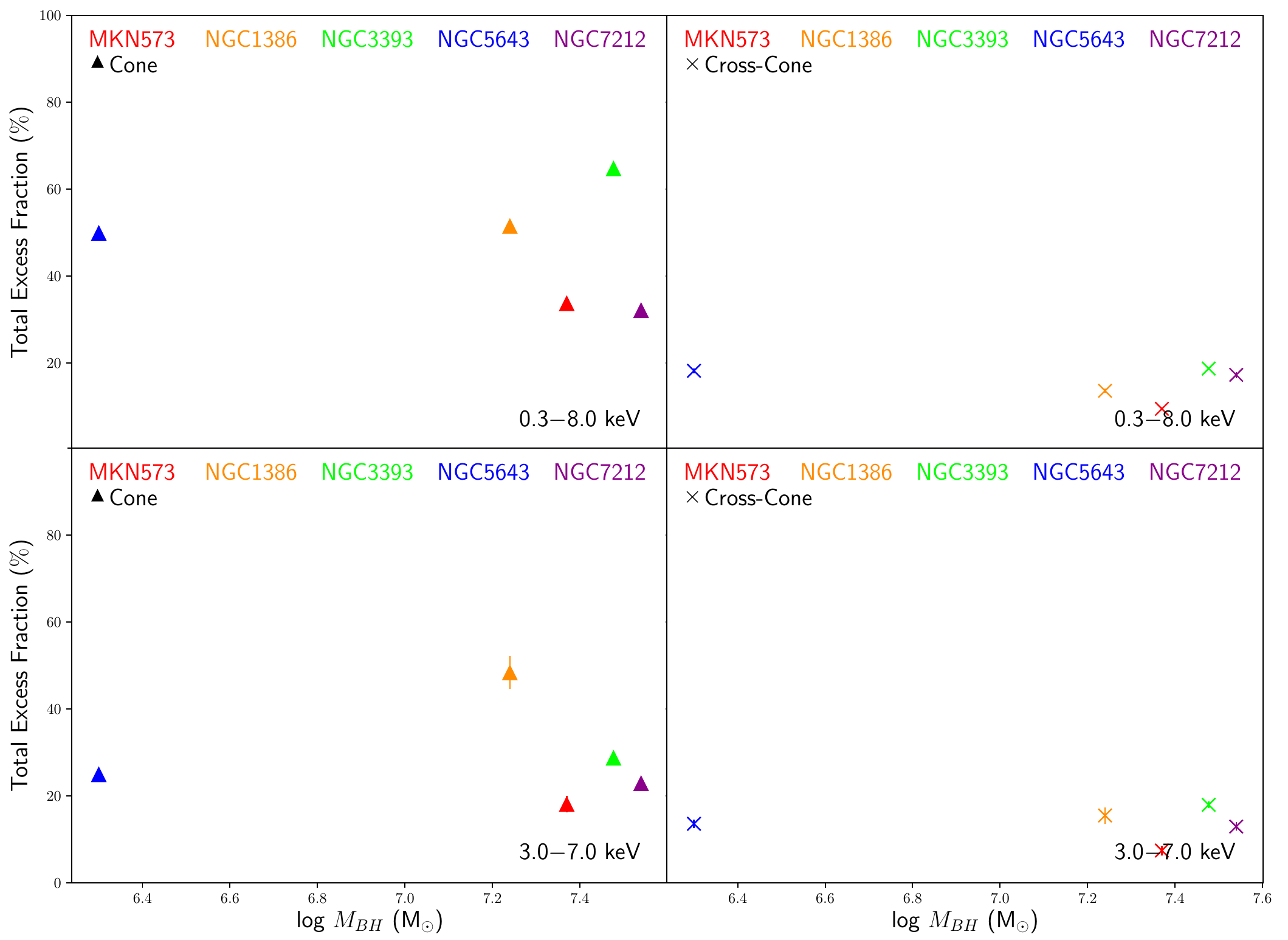}} \\
\end{tabular}
\caption{Total excess fraction in the cone (left) and cross-cone (right) regions for each of the five CT AGN as a function of black hole mass ($\log M_{BH}$; Table \ref{tab:prop}) for both the wide-band ($0.3-8.0$ keV; top) and hard-band ($3.0-7.0$ keV; bottom). There is not much evidence that $M_{BH}$, influences the total excess fraction in either energy band, but the cone region is more dispersed than the cross-cone region.
\label{fig:fex:mbh}}
\end{center}
\end{figure*}
%%%%%%%%%%%%%%%%%%%%%

The black hole masses for our sources are calculated from independent dynamical measurements (\citealt{Bia07,Her15} and determined from water maser observations (NGC 3393; \citealt{Kon08}); Table \ref{tab:prop}). While the masses are not well constrained, we do not see a strong dependence on the total excess fraction with black hole mass for either energy band or region (Figure \ref{fig:fex:mbh}). The Pearson coefficients of these dependences are -0.18 (cone) and -0.26 (cross-cone) in $0.3-8.0$ keV, and 0.17 (cone) and 0.003 (cross-cone) in $3.0-7.0$ keV.

\subsubsection{``Extended'' X-ray Luminosity}

%%%%%%%%%%%%%%%%%%%%%
% FIGURE : FEX PROPERTIES : Lext
\begin{figure*}
\begin{center}
\begin{tabular}{cc}
\resizebox{67mm}{!}{\includegraphics{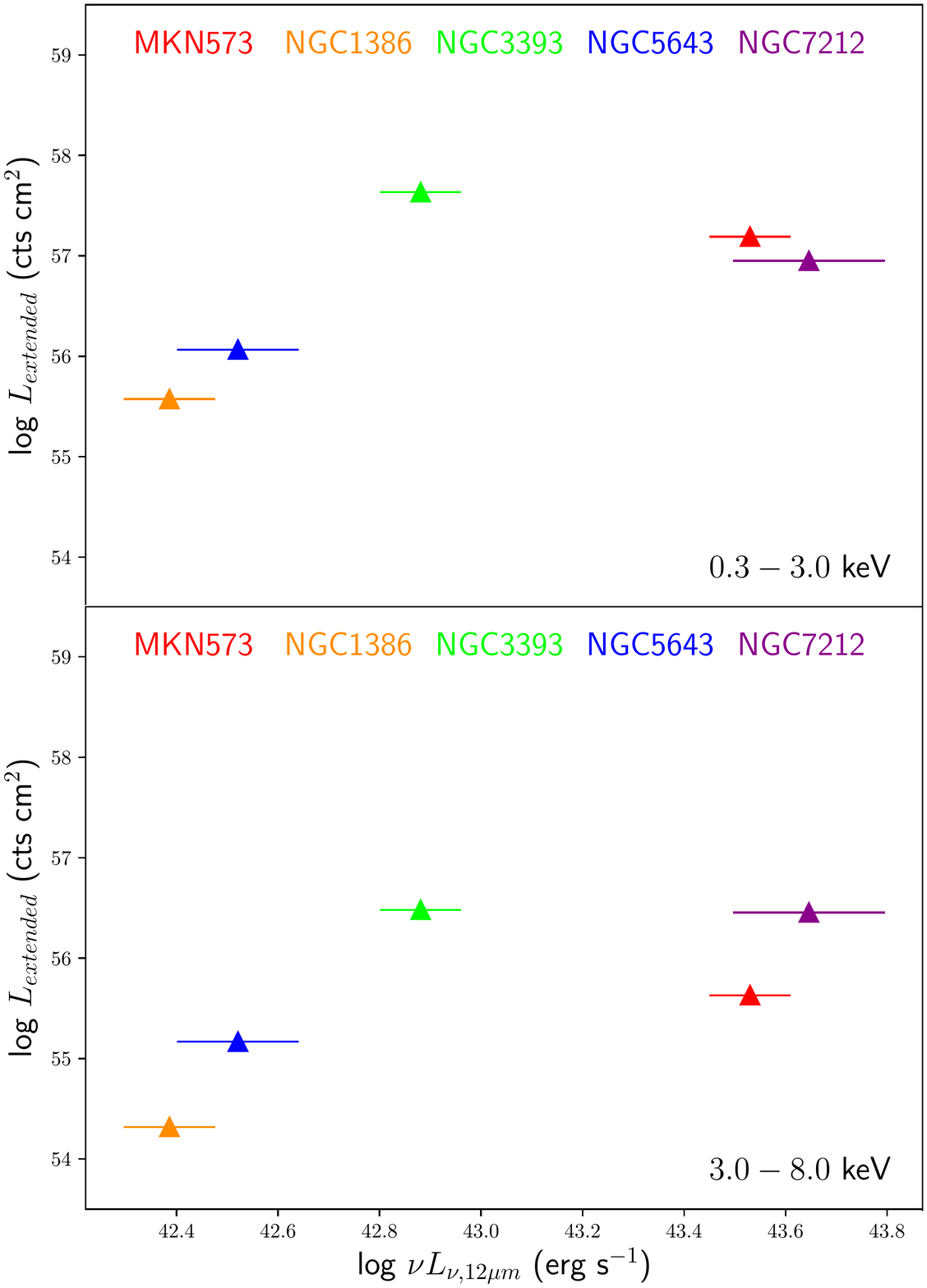}} \\
\end{tabular}
\caption{Luminosity of the extended emission in the cone (left) and cross-cone (right) regions for each of the five CT AGN as a function of the nuclear luminosity for both the soft energy band ($0.3-3.0$ keV; top) and hard-band ($3.0-8.0$ keV; bottom). 
We use the distance and PSF corrected, extended counts (Tables \ref{tab:cts:573},\ref{tab:cts:1386},\ref{tab:cts:3393},\ref{tab:cts:5643},\ref{tab:cts:7212}) as a proxy for the extended luminosity and the 12 micron luminosity as a proxy for the nuclear luminosity (log $\nu L_{\nu,12\mu m}$; Table \ref{tab:prop}). 
We find enhanced extended luminosities at higher nuclear luminosities, likely due to the increased availability of nuclear photons. \label{fig:lex}}
\end{center}
\end{figure*}
%%%%%%%%%%%%%%%%%%%%%

We then explore whether there is a correlation between the nuclear luminosity (for which we use $12\mu m L_{IR}$ as a proxy) with the luminosity of the extended emission (Figure \ref{fig:lex}). We calculate the extended luminosity using the extended counts (PSF subtracted counts $>1.5$\arcsec; Tables \ref{tab:cts:573},\ref{tab:cts:1386},\ref{tab:cts:3393},\ref{tab:cts:5643},\ref{tab:cts:7212}) multiplied by $4\pi d^2$, where \textit{d} is the distance in cm. We find a positive correlation between the extended luminosity with the nuclear luminosity for both the soft ($0.3-3.0$ keV; Pearson coefficient of 0.72) and hard ($3.0-8.0$ keV; Pearson coefficient of 0.72) X-ray bands. This is unsurprising, because with higher nuclear luminosities, there are more photons available to interact with the surrounding medium, boosting the extended luminosities.

%%%%%%%%%%%%%%%%%%%%%%%%%%%%%%%%%%%%%%%%%%%%%%%%%
% CONCLUSION
%%%%%%%%%%%%%%%%%%%%%%%%%%%%%%%%%%%%%%%%%%%%%%%%%
\section{Conclusion}\label{sec:con}

We have analyzed and compared the spatial extent of five CT AGN that exhibit extended X-ray emission in \textit{Chandra} observations .

\begin{enumerate}

\item We find extended emission in the cone region for all five CT AGN in our sample, MKN 573, NGC 1386, NGC 3393, NGC 5643, and NGC 7212. This emission is significant across all three of our metrics, including, counts over the PSF, the total excess fraction, and the extended fraction for both the full X-ray band ($0.3-8.0$ keV) and surrounding the Fe K$\alpha$ line ($6-7$ keV).

\item We find extended emission in the cross-cone region for the majority of our CT AGN. We find significant counts over the \textit{Chandra} PSF, total excess fraction, and extended fraction for the entire sample from $0.3-8.0$ keV. For $6-7$ keV, however, MKN 573 and NGC 1386 do not exhibit $>3\sigma$ significance in the extended fraction (although counts $>0.5$\arcsec\, and total excess fraction are significant).

\item We show that the extent as a function of energy at 1\% of the surface brightness in the cone region of our CT AGN exhibits a steeper relationship compared to the cross-cone region (with the exception of NGC 3393 that exhibits a similar cone-cross-cone slope). This may be explained by an inclination effect, in which the orientation of the ionization cone with respect to the galactic disk molecular clouds impacts the extent to which the soft X-rays propagate through the ISM. We further investigate these slopes as a function of galaxy properties and find for the cone region that the slope decreases as the mass of the black hole increases.

\item We compare the extent of our sample at 1\% of the surface brightness with the column densities for both the cone and cross-cone region. In the cone region, we find a positive trend between $N_H$ and the X-ray extent that may indicate that nuclear obscuration along the ionization cone is correlated with the disk molecular clouds. The cross-cone does not exhibit a trend, suggesting that the nuclear torus is the dominant obscurer.

\item We do not find a clear correlation between the excess fraction and black hole mass for our sample, despite finding that as the black hole mass increases, the soft X-rays extend farther than the hard X-rays.
We do find, however, a shallow trend in the excess fraction of our five CT AGN in the cone region with host galaxy extent (d$_{25}$), but no trend in the cross-cone region.

\item We use $L_{IR}$ as a proxy for the nuclear luminosity and investigate how luminosity impacts the luminosity of the extended X-rays. To do this, we use the $counts*4\pi d^2$ as a proxy for the extended luminosity. We find that as the nuclear luminosity increases, the extended luminosity also increases, likely due to the heightened availability of AGN photons.

\end{enumerate}

%%%%%%%%%%%%%%%%%%%%%%%%%%%%%%%%%%%%%%%%%%%%%%%%%
\acknowledgments

We would like to thank the referee for constructive comments and suggestions to improve this paper. The analysis of NGC 3393 was in part, the Master thesis of K. Parker at University of Southampton, UK. This work makes use of data from the \textit{Chandra} data archive, and the NASA-IPAC Extragalactic Database (NED). The analysis makes use of CIAO and Sherpa, developed by the \textit{Chandra} X-ray Center; SAOImage ds9; XSPEC, developed by HEASARC at NASA-GSFC; and the Astrophysics Data System (ADS). JW acknowledges support by the National Key R\&D Program of China (2016YFA0400702) and the NSFC grants (U1831205, 11473021, 11522323). This work was supported by the Chandra Guest Observer programs, grant no. GO5-16101X
GO7-18112X, GO8-19099X, GO8-19096X (PI: Maksym); GO8$-$19074X (PI: Fabbiano).

%%%%%%%%%%%%%%%%%%%%%%%%%%%%%%%%%%%%%%%%%%%%%%%%%
\begin{appendix}

\section{X-ray Extent as a Function of Energy}

\begin{figure*}
\begin{center}\footnotesize
\begin{tabular}{cc}
\resizebox{160mm}{!}{\includegraphics{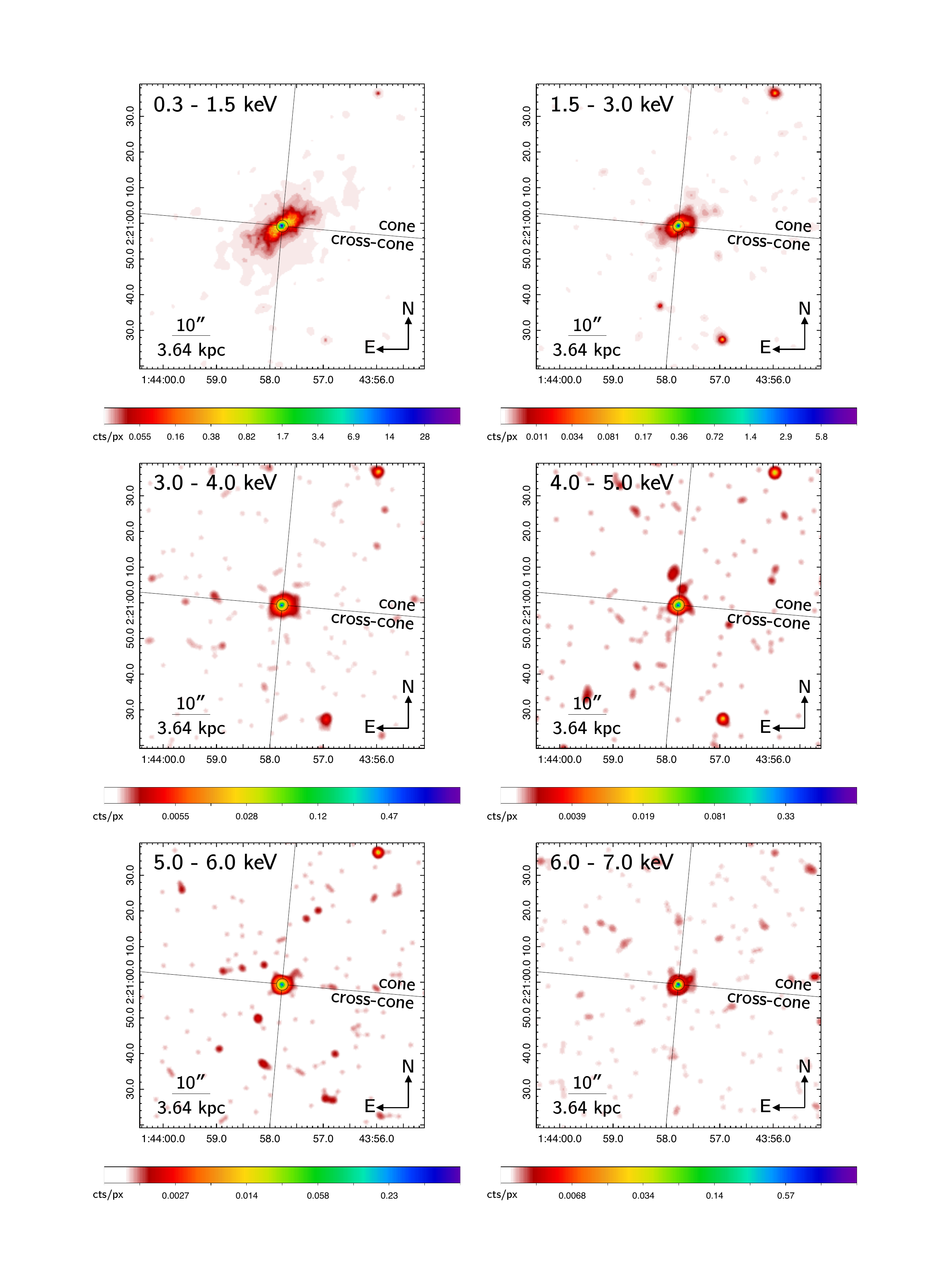}} \\
\end{tabular}
\caption{Adaptively smoothed images of MKN 573 in the indicated energy bands on image pixel = 1/8 ACIS pixel (\textit{dmingadapt}; $0.5-15$ pixel scales, $2-5$ counts under kernel, 30 iterations). The image contours are logarithmic with colors corresponding to the number of counts per image pixel. The box size is 80\arcsec\, $\times$ 80\arcsec. Also shown are the 1.5\arcsec\,(0.546 kpc) circular region and cone/cross-cone regions. \label{fig:grid:573}}
\end{center}
\end{figure*}

\begin{figure*}
\begin{center}\footnotesize
\begin{tabular}{cc}
\resizebox{160mm}{!}{\includegraphics{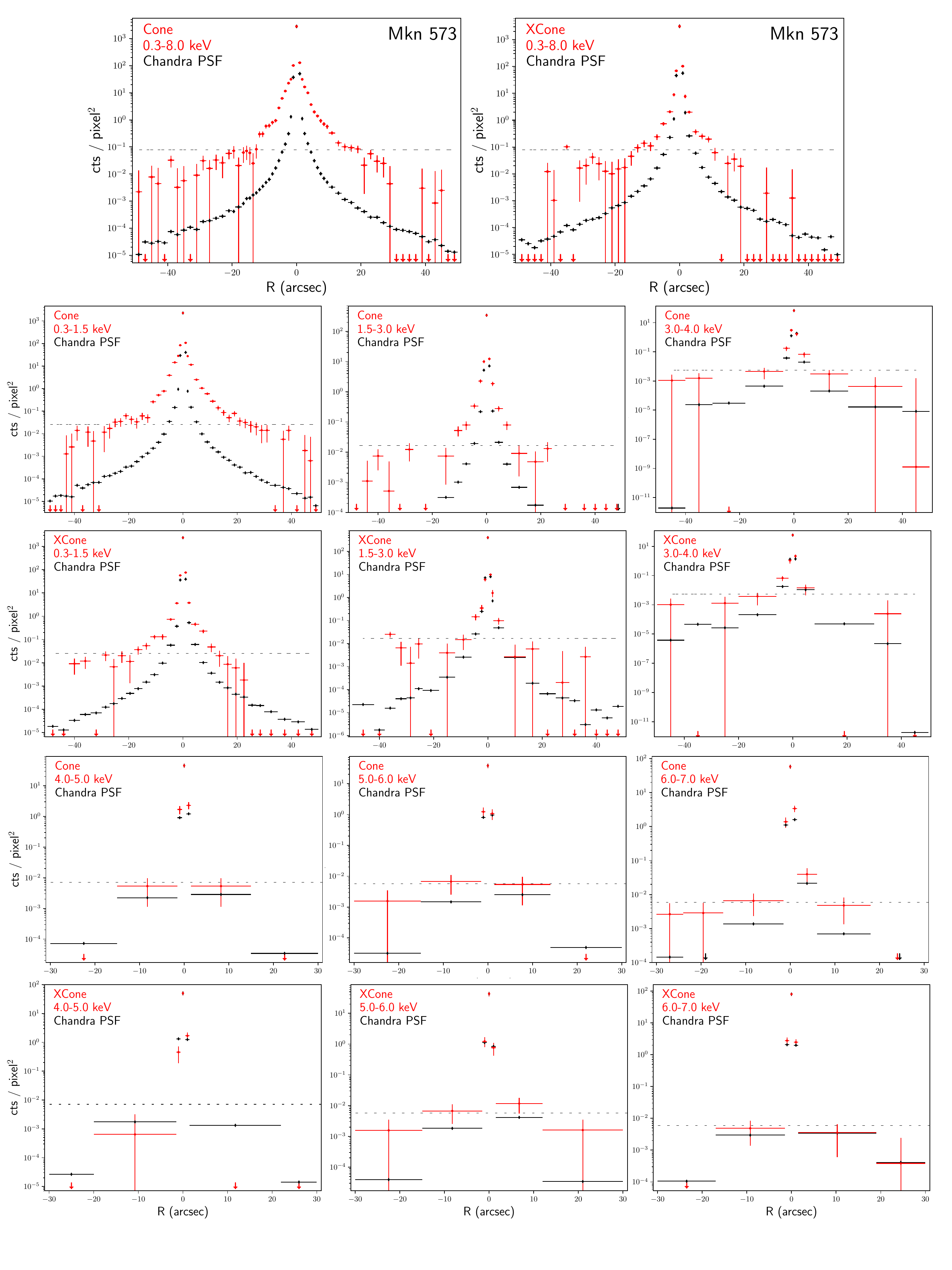}} \\
\end{tabular}
\caption{Background subtracted radial profiles of MKN 573 for the indicated energy bands compared to the \textit{Chandra} PSF which has been re-normalized to the 0.5\arcsec~ (0.182 kpc) nuclear region for both the cone and cross-cone regions (as labeled ``Cone'' and ``XCone'', respectively). Each bin contains a minimum of 10 counts and is shown with $1\sigma$ error. We include a dashed horizontal line to indicate the level of background emission and note that points below this line are valid data since the background has already been subtracted. Downward arrows indicate a region at or below the background. \label{fig:rp:573}}
\end{center}
\end{figure*}

\begin{figure*}
\begin{center}\footnotesize
\begin{tabular}{cc}
\resizebox{160mm}{!}{\includegraphics{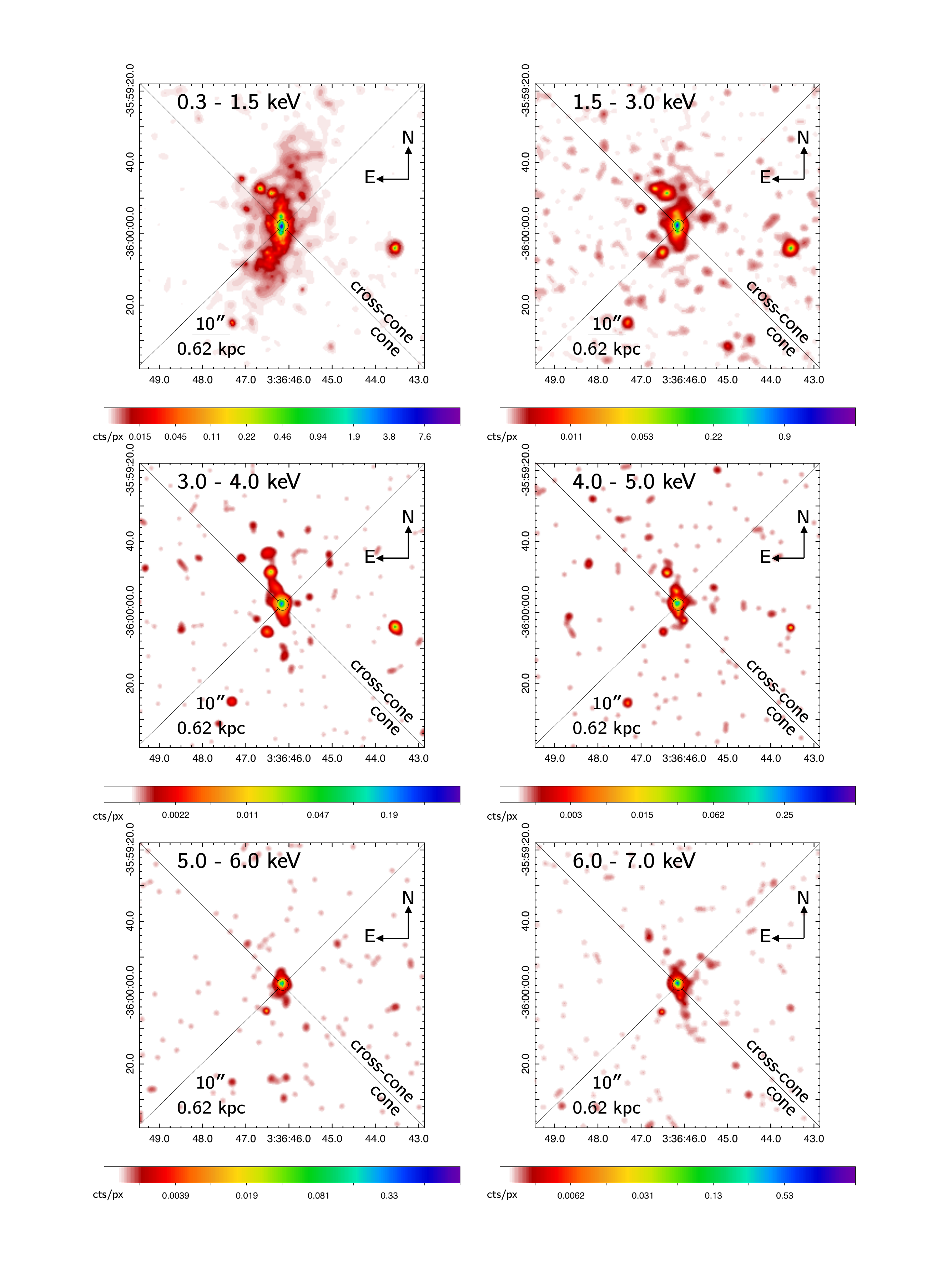}} \\
\end{tabular}
\caption{Adaptively smoothed images of NGC 1386 in the indicated energy bands on image pixel = 1/8 ACIS pixel (\textit{dmingadapt}; $0.5-15$ pixel scales, $2-5$ counts under kernel, 30 iterations). The image contours are logarithmic with colors corresponding to the number of counts per image pixel. The box size is 80\arcsec\, $\times$ 80\arcsec. Also shown are the 1.5\arcsec\,(0.093 kpc) circular region and cone/cross-cone regions. \label{fig:grid:1386}}
\end{center}
\end{figure*}

\begin{figure*}
\begin{center}\footnotesize
\begin{tabular}{cc}
\resizebox{160mm}{!}{\includegraphics{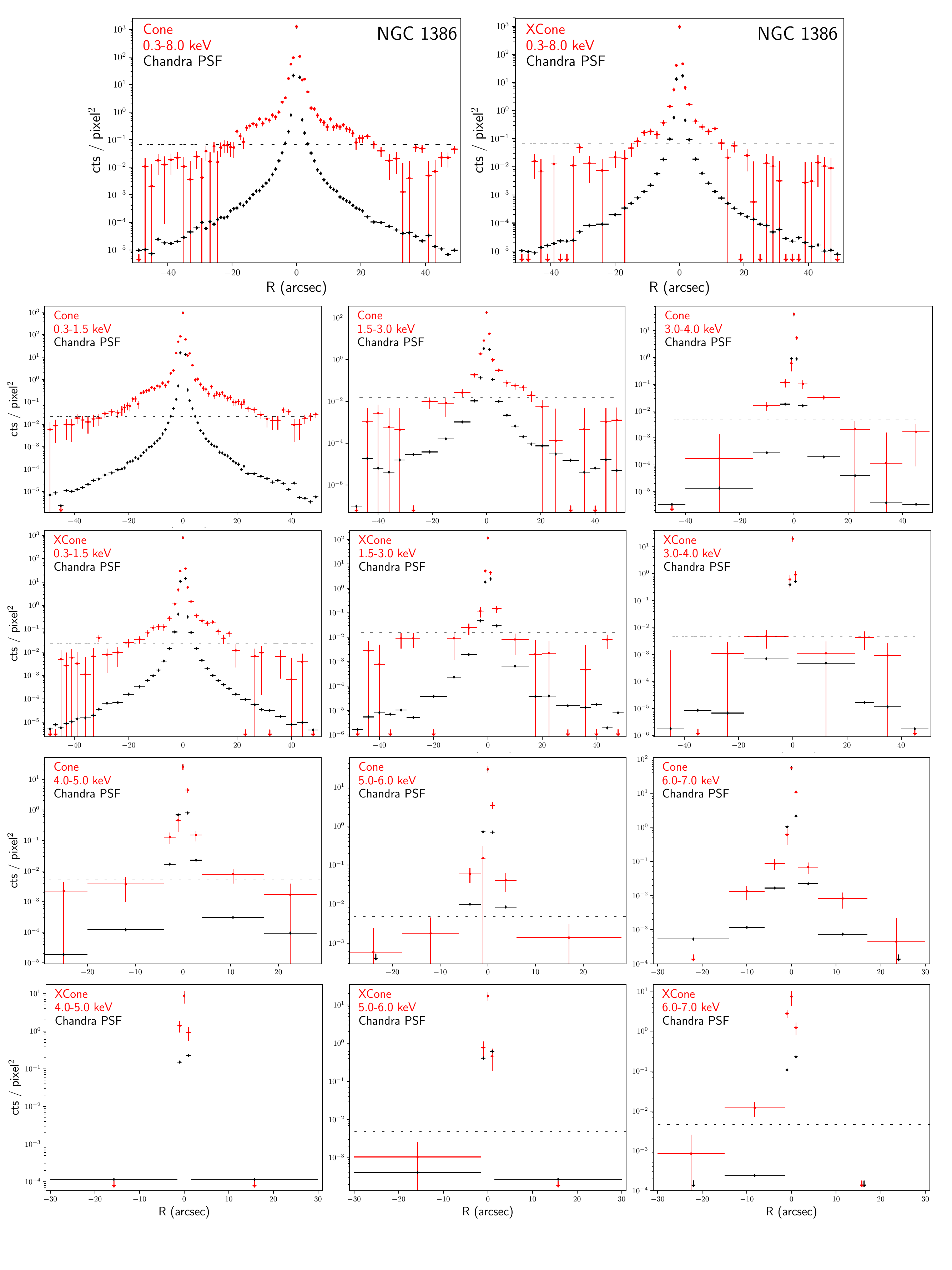}} \\
\end{tabular}
\caption{Background subtracted radial profiles of NGC 1386 for the indicated energy bands compared to the \textit{Chandra} PSF which has been re-normalized to the 0.5\arcsec~ (0.031 kpc) nuclear region for both the cone and cross-cone regions (as labeled ``Cone'' and ``XCone'', respectively). Each bin contains a minimum of 10 counts and is shown with $1\sigma$ error. We include a dashed horizontal line to indicate the level of background emission and note that points below this line are valid data since the background has already been subtracted. Downward arrows indicate a region at or below the background. \label{fig:rp:1386}}
\end{center}
\end{figure*}

\begin{figure*}
\begin{center}\footnotesize
\begin{tabular}{cc}
\resizebox{160mm}{!}{\includegraphics{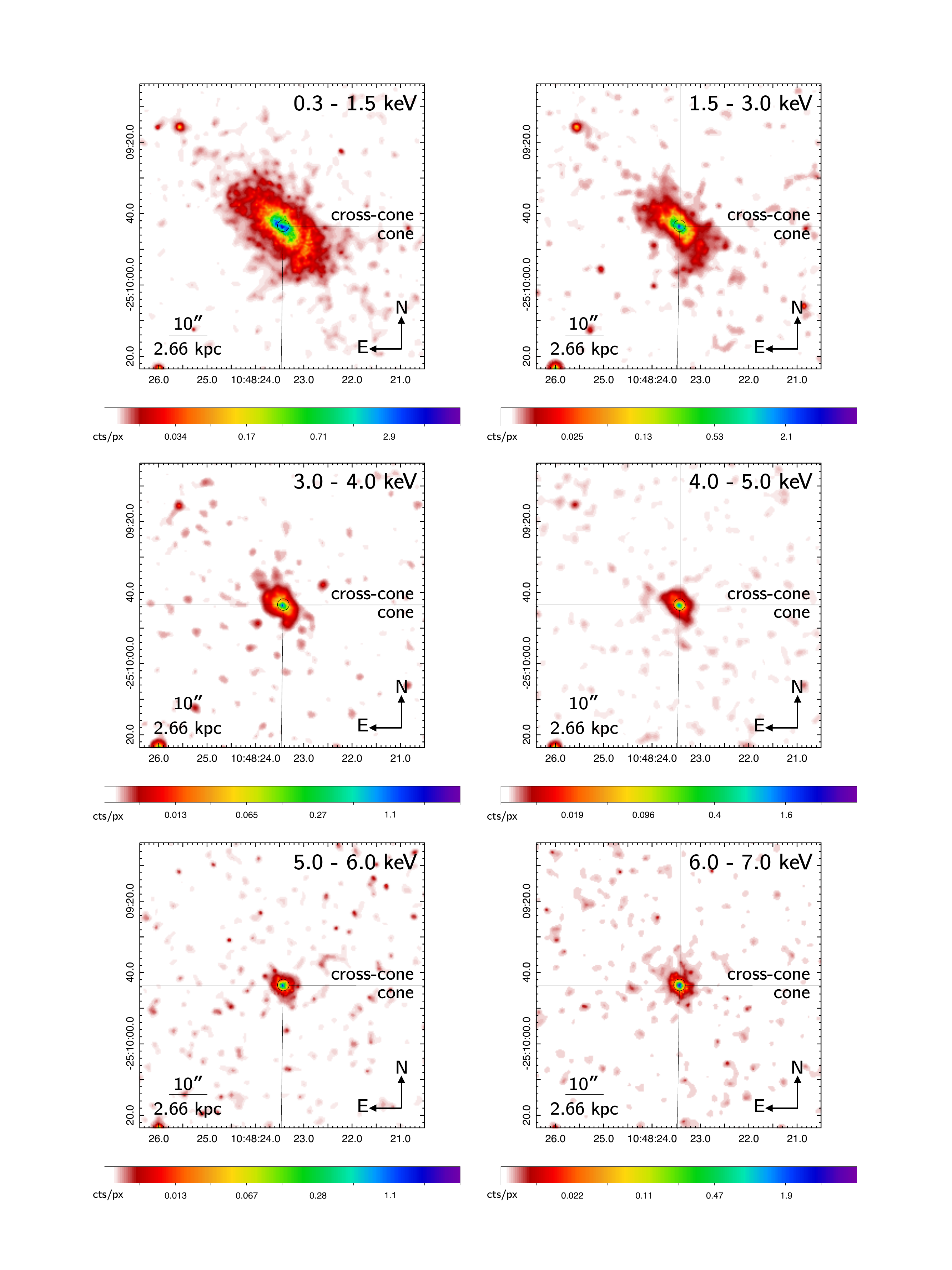}} \\
\end{tabular}
\caption{Adaptively smoothed images of NGC 3393 in the indicated energy bands on image pixel = 1/8 ACIS pixel (\textit{dmingadapt}; $0.5-15$ pixel scales, $2-5$ counts under kernel, 30 iterations). The image contours are logarithmic with colors corresponding to the number of counts per image pixel. The box size is 80\arcsec\, $\times$ 80\arcsec. Also shown are the 1.5\arcsec\,(0.399 kpc) circular region and cone/cross-cone regions. \label{fig:grid:3393}}
\end{center}
\end{figure*}

\begin{figure*}
\begin{center}\footnotesize
\begin{tabular}{cc}
\resizebox{160mm}{!}{\includegraphics{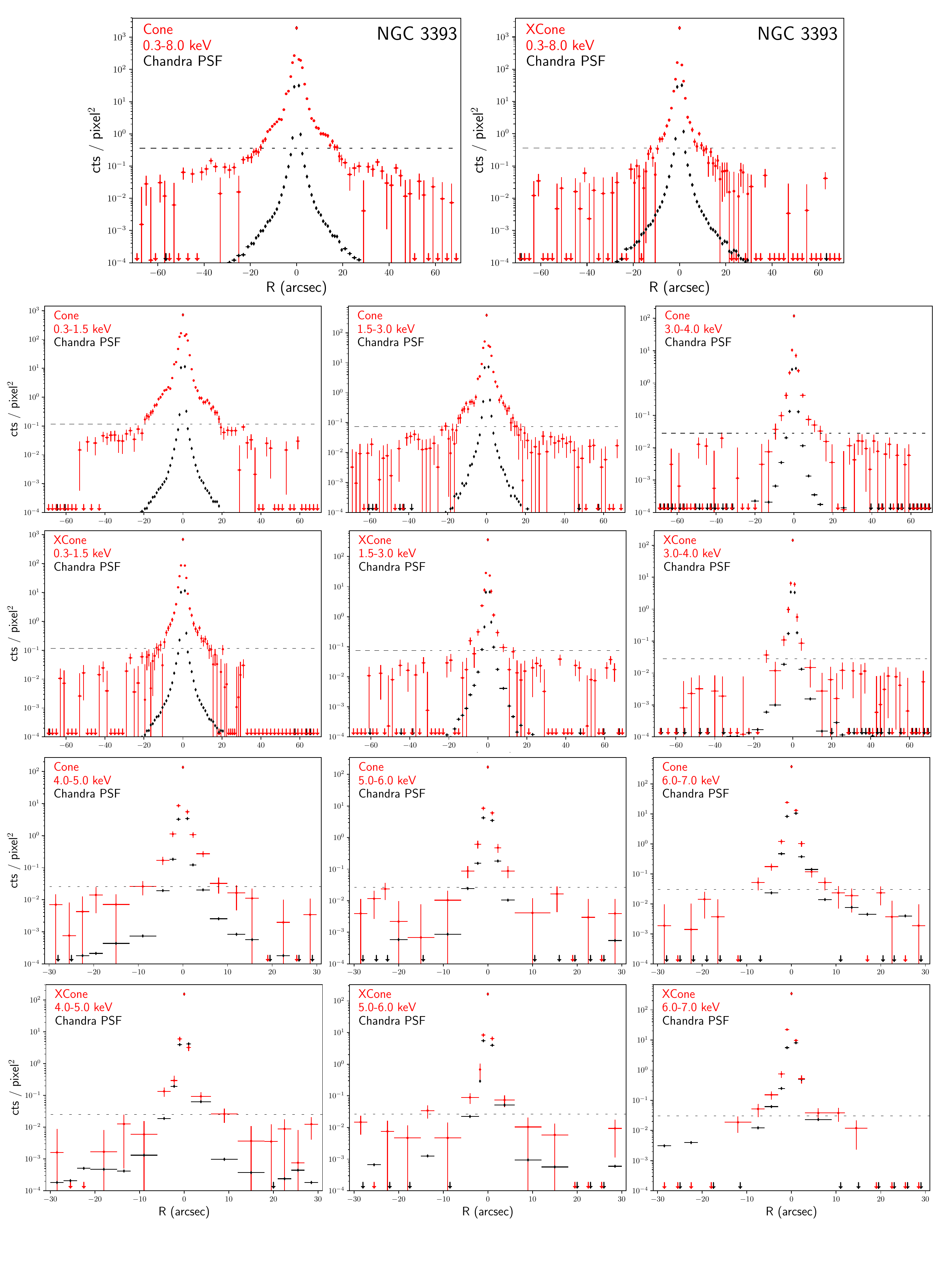}} \\
\end{tabular}
\caption{Background subtracted radial profiles of NGC 3393 for the indicated energy bands compared to the \textit{Chandra} PSF which has been re-normalized to the 0.5\arcsec~ (0.133 kpc) nuclear region for both the cone and cross-cone regions (as labeled ``Cone'' and ``XCone'', respectively). Each bin contains a minimum of 10 counts and is shown with $1\sigma$ error. We include a dashed horizontal line to indicate the level of background emission and note that points below this line are valid data since the background has already been subtracted. Downward arrows indicate a region at or below the background. \label{fig:rp:3393}}
\end{center}
\end{figure*}

\begin{figure*}
\begin{center}\footnotesize
\begin{tabular}{cc}
\resizebox{160mm}{!}{\includegraphics{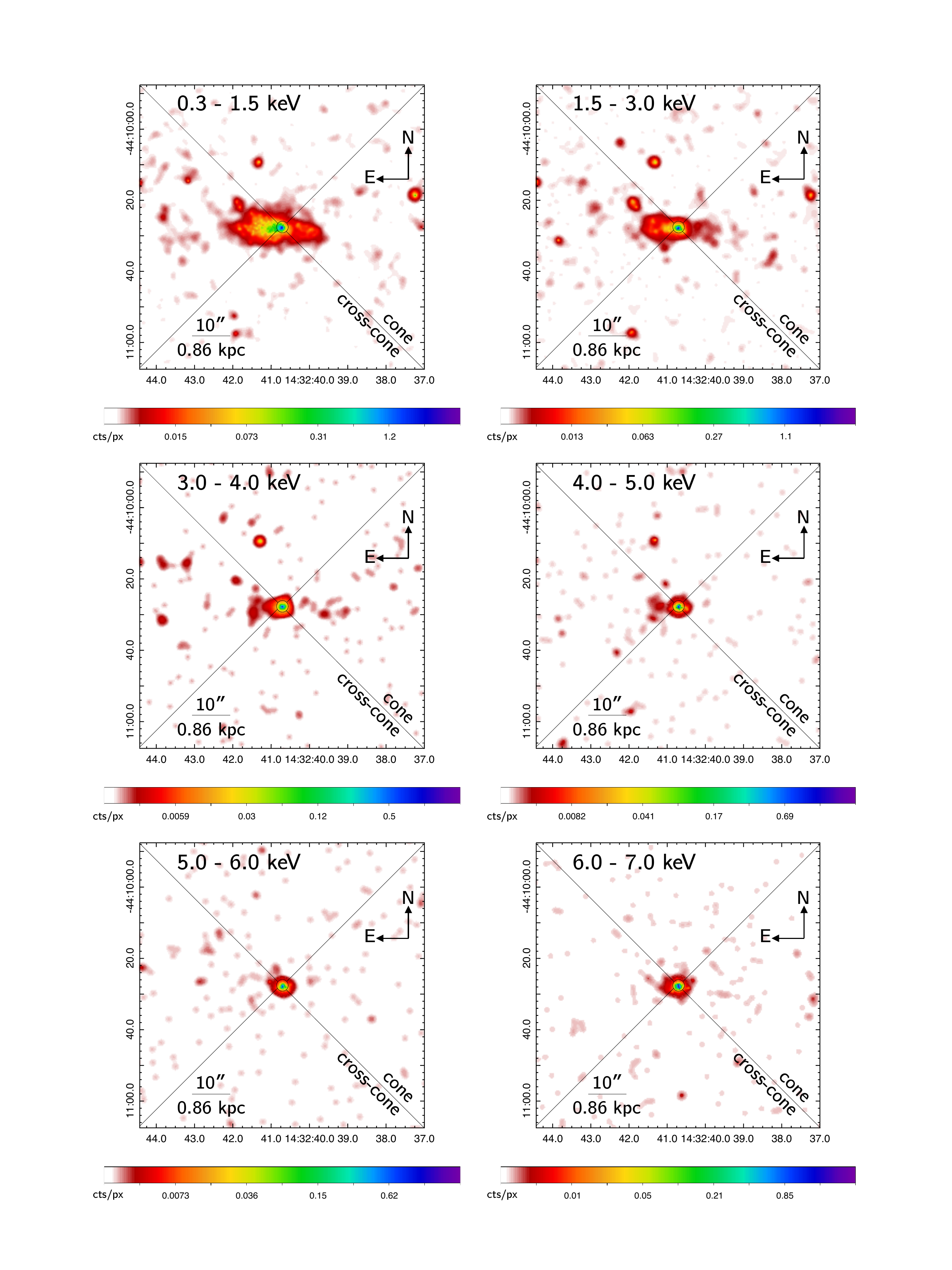}} \\
\end{tabular}
\caption{Adaptively smoothed images of NGC 5643 in the indicated energy bands on image pixel = 1/8 ACIS pixel (\textit{dmingadapt}; $0.5-15$ pixel scales, $2-5$ counts under kernel, 30 iterations). The image contours are logarithmic with colors corresponding to the number of counts per image pixel. The box size is 80\arcsec\, $\times$ 80\arcsec. Also shown are the 1.5\arcsec\,(0.129 kpc) circular region and cone/cross-cone regions. \label{fig:grid:5643}}
\end{center}
\end{figure*}

\begin{figure*}
\begin{center}\footnotesize
\begin{tabular}{cc}
\resizebox{160mm}{!}{\includegraphics{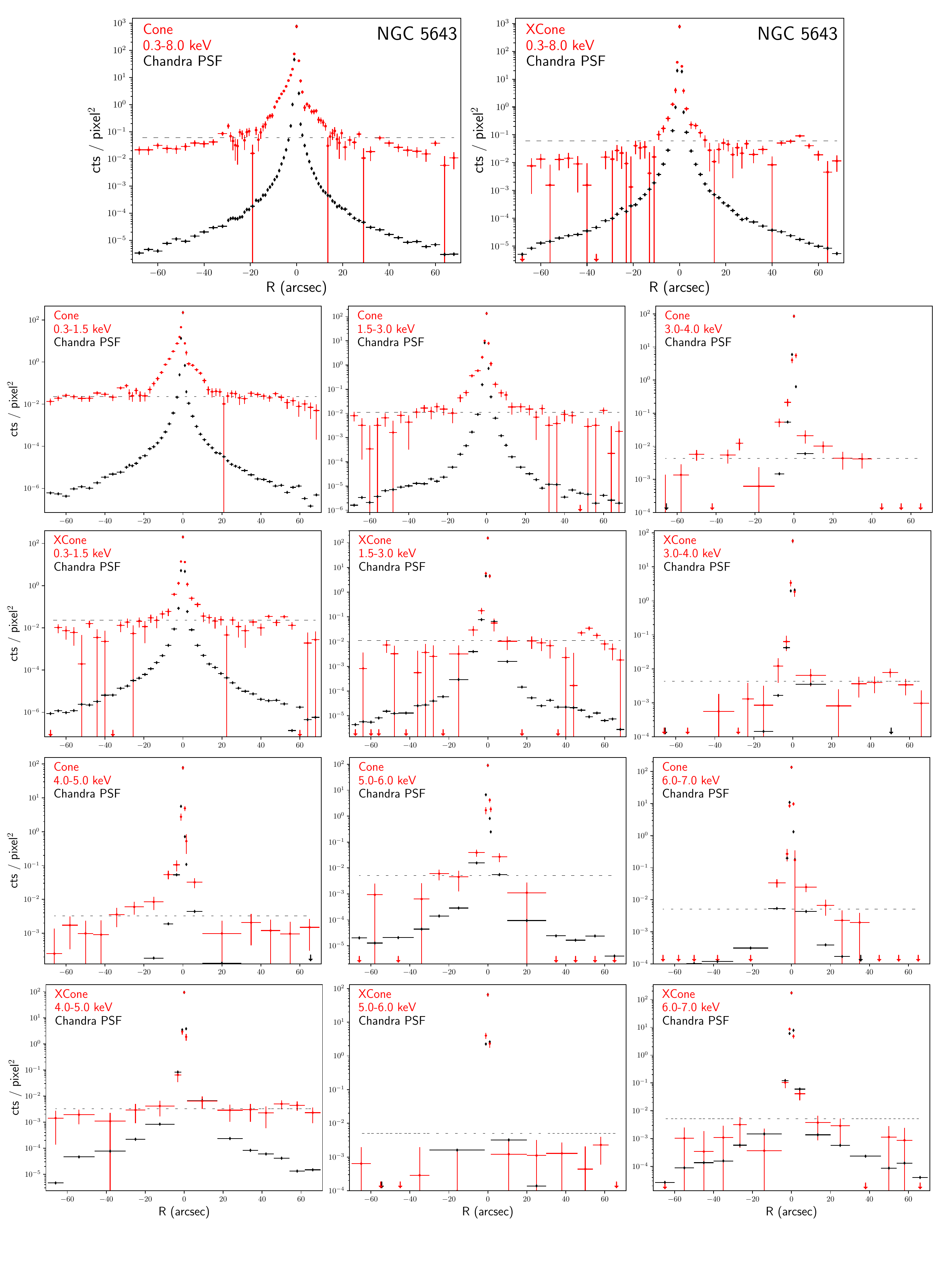}} \\
\end{tabular}
\caption{Background subtracted radial profiles of NGC 5643 for the indicated energy bands compared to the \textit{Chandra} PSF which has been re-normalized to the 0.5\arcsec~ (0.043 kpc) nuclear region for both the cone and cross-cone regions (as labeled ``Cone'' and ``XCone'', respectively). Each bin contains a minimum of 10 counts and is shown with $1\sigma$ error. We include a dashed horizontal line to indicate the level of background emission and note that points below this line are valid data since the background has already been subtracted. Downward arrows indicate a region at or below the background. \label{fig:rp:5643}}
\end{center}
\end{figure*}

\begin{figure*}
\begin{center}\footnotesize
\begin{tabular}{cc}
\resizebox{160mm}{!}{\includegraphics{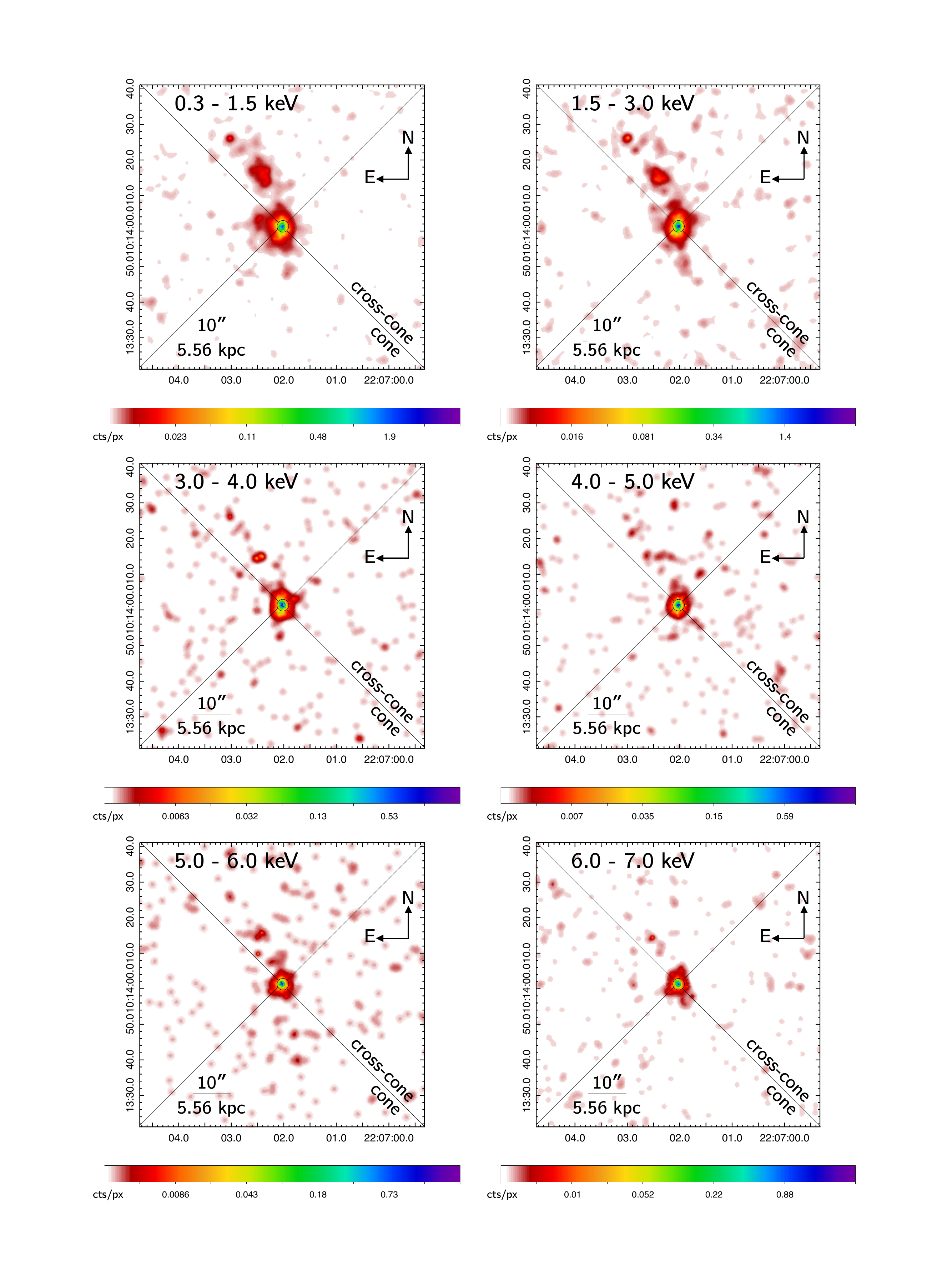}} \\
\end{tabular}
\caption{Adaptively smoothed images of NGC 7212 in the indicated energy bands on image pixel = 1/8 ACIS pixel (\textit{dmingadapt}; $0.5-15$ pixel scales, $2-5$ counts under kernel, 30 iterations). The image contours are logarithmic with colors corresponding to the number of counts per image pixel. The box size is 80\arcsec\, $\times$ 80\arcsec. Also shown are the 1.5\arcsec\,(0.834 kpc) circular region and cone/cross-cone regions. \label{fig:grid:7212}}
\end{center}
\end{figure*}

\begin{figure*}
\begin{center}\footnotesize
\begin{tabular}{cc}
\resizebox{160mm}{!}{\includegraphics{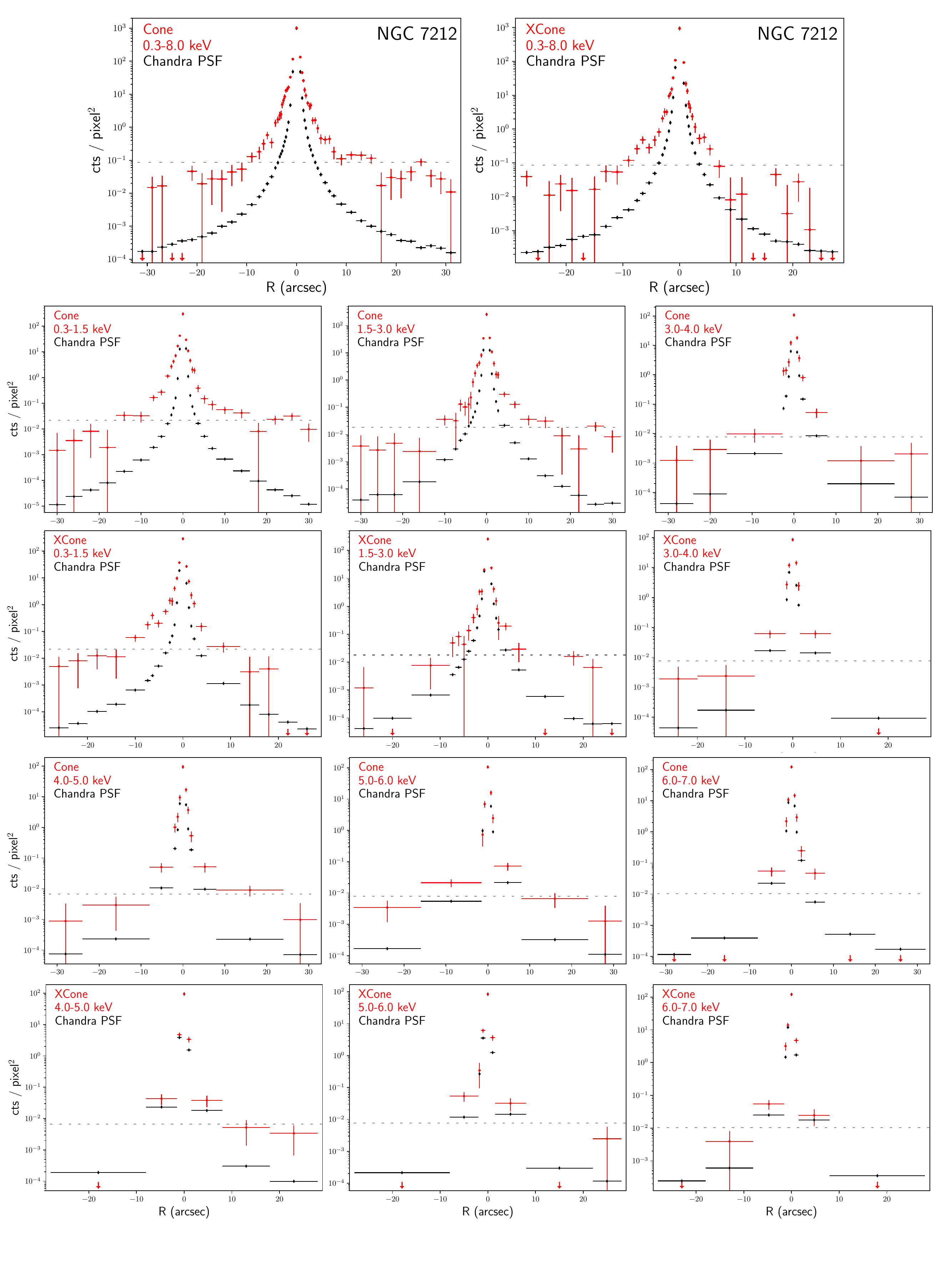}} \\
\end{tabular}
\caption{Background subtracted radial profiles of NGC 7212 for the indicated energy bands compared to the \textit{Chandra} PSF which has been re-normalized to the 0.5\arcsec~ (0.278 kpc) nuclear region for both the cone and cross-cone regions (as labeled ``Cone'' and ``XCone'', respectively). Each bin contains a minimum of 10 counts and is shown with $1\sigma$ error. We include a dashed horizontal line to indicate the level of background emission and note that points below this line are valid data since the background has already been subtracted. Downward arrows indicate a region at or below the background. \label{fig:rp:7212}}
\end{center}
\end{figure*}

We have appended the adaptively smoothed images of our five CT AGN, each divided into six energy bins between $0.3-7.0$ keV: MKN 573 (Figure \ref{fig:grid:573}), NGC 1386 (Figure \ref{fig:grid:1386}), NGC 3393 (Figure \ref{fig:grid:3393}), NGC 5643 (Figure \ref{fig:grid:5643}), NGC 7212 (Figure \ref{fig:grid:7212}). 
We have also included the radial profiles of our five CT AGN for both the cone and cross-cone regions that correspond to the energy bins in the adaptively smoothed images: MKN 573 (Figure \ref{fig:rp:573}), NGC 1386 (Figure \ref{fig:rp:1386}), NGC 3393 (Figure \ref{fig:rp:3393}), NGC 5643 (Figure \ref{fig:rp:5643}), NGC 7212 (Figure \ref{fig:rp:7212}).
We fit these radial profiles using the spline approximation described in Section \ref{sec:rp} and calculate the FWHM and extent at 1\% of the surface brightness for both the cone and cross-cone region and each of the energy bins.

%%%%%%%%%%%%%%%%%%%%%
% FIGURE : IMAGES
%%%%%%%%%%%%%%%%%%%%%

%%%%%%%%%%%%%%%%%%%%%
% FIGURE : RADIAL PROFILES
%%%%%%%%%%%%%%%%%%%%%

%%%%%%%%%%%%%%%%%%%%%
% FIGURE : IMAGES
%%%%%%%%%%%%%%%%%%%%%

%%%%%%%%%%%%%%%%%%%%%
% FIGURE : RADIAL PROFILES
%%%%%%%%%%%%%%%%%%%%%

%%%%%%%%%%%%%%%%%%%%%
% FIGURE : IMAGES
%%%%%%%%%%%%%%%%%%%%%

%%%%%%%%%%%%%%%%%%%%%
% FIGURE : RADIAL PROFILES
%%%%%%%%%%%%%%%%%%%%%

%%%%%%%%%%%%%%%%%%%%%
% FIGURE : IMAGES
%%%%%%%%%%%%%%%%%%%%%

%%%%%%%%%%%%%%%%%%%%%
% FIGURE : RADIAL PROFILES
%%%%%%%%%%%%%%%%%%%%%

%%%%%%%%%%%%%%%%%%%%%
% FIGURE : IMAGES
%%%%%%%%%%%%%%%%%%%%%

%%%%%%%%%%%%%%%%%%%%%
% FIGURE : RADIAL PROFILES
%%%%%%%%%%%%%%%%%%%%%

\end{appendix}

%%%%%%%%%%%%%%%%%%%%%%%%%%%%%%%%%%%%%%%%%%%%%%%%%
\clearpage
\bibliography{ehx}

%%%%%%%%%%%%%%%%%%%%%%%%%%%%%%%%%%%%%%%%%%%%%%%%%

\end{document}